\documentclass[aip,jcp,amsmath,amssymb,longbibliography,reprint]{revtex4-1}

\usepackage[utf8]{inputenc}
\usepackage[T1]{fontenc}
\usepackage{mathptmx}
\usepackage{etoolbox}
\usepackage{graphicx}
\usepackage[hidelinks]{hyperref}
\usepackage{bm}
\usepackage{color}

\makeatletter
\def\@email#1#2{%
 \endgroup
 \patchcmd{\titleblock@produce}
  {\frontmatter@RRAPformat}
  {\frontmatter@RRAPformat{\produce@RRAP{*#1\href{mailto:#2}{#2}}}\frontmatter@RRAPformat}
  {}{}
}%
\makeatother

\newcommand{\rv}{{\mathbf r}}

\renewcommand{\vec}{\mathbf}

\newcommand{\eqr}[1]{Eq.~\eqref{#1}}

\newcommand{\mydelete}[1]{{}}

\newcommand{\rmexc}{{\rm exc}}
\newcommand{\rmext}{{\rm ext}}

\begin{document}

\title{Learning the bulk and interfacial physics of liquid-liquid phase separation with neural density functionals}

\author{Silas Robitschko}
\affiliation{Theoretische Physik II, Physikalisches Institut, Universität Bayreuth, D-95447 Bayreuth, Germany}
\author{Florian Sammüller}
\affiliation{Theoretische Physik II, Physikalisches Institut, Universität Bayreuth, D-95447 Bayreuth, Germany}
\author{Matthias Schmidt}
\affiliation{Theoretische Physik II, Physikalisches Institut, Universität Bayreuth, D-95447 Bayreuth, Germany}
\email{Matthias.Schmidt@uni-bayreuth.de}
\author{Robert Evans}
\affiliation{H.~H.~Wills Physics Laboratory, University of Bristol, Royal Fort, Bristol BS8 1TL, United Kingdom}

\date{\today}

\begin{abstract}
We use simulation-based supervised machine learning and classical density functional theory to investigate bulk and interfacial phenomena associated with phase coexistence in binary mixtures.
For a prototypical symmetrical Lennard-Jones mixture our trained neural density functional yields accurate liquid-liquid and liquid-vapour binodals together with predictions for the variation of the associated interfacial tensions across the entire fluid phase diagram.
From the latter we determine the contact angles at fluid-fluid interfaces along the line of triple-phase coexistence and confirm there can be no wetting transition in this symmetrical mixture.
\end{abstract}

\maketitle

Making accurate predictions for the phase behaviour of complex systems remains a major computational challenge, despite the availability of a wide variety of flexible simulation methodology \cite{frenkel2023book}.
The particular phenomenon of liquid-liquid phase separation occurs across a broad spectrum of substances, from mixtures of simple (rare gas) liquids \cite{vanKonynenburg1980}, to models of water \cite{sciortino2025} and ouzo \cite{archer2024ouzo,sibley2025}.
Moreover, liquid-liquid phase separation is argued to be a possible structure formation mechanism in biological cells \cite{hyman2014}, where much theoretical work has been carried out on the basis of the Flory-Huggins model.
Studying the emergence of phase transitions provides fertile ground for the development of machine-learning strategies \cite{arnold2024, bedolla2021, chertenkov2023} and for realizing inverse design of soft matter \cite{dijkstra2021ml, coli2022scienceAdvances}.
Identifying signs of critical behaviour is also important when addressing dynamical questions \cite{liu2024, harris2024, morr2024}.

Classical density functional theory (DFT) \cite{evans1979, evans1992, evans2016} provides a suitable theoretical framework in which machine learning can be integrated naturally, as was shown in early work \cite{teixera2014, lin2019ml, lin2020ml, cats2022ml, yatsyshin2022, fang2022, qiao2020}, for anisotropic particles \cite{simon2023mlPatchy, simon2024patchy, yang2024} and in a variety of further contexts \cite{glitsch2025disks, dijkman2024ml, kelley2024ml, malpica-morales2023, pan2025}.
Bui and Cox have used machine learning to address solvation across length scales \cite{bui2024}, electromechanics \cite{bui2025electromechanics}, dielectrocapillarity \cite{bui2025dielectrocapillarity} and ionic fluids \cite{bui2024neuralrpm}.
In this paper we build upon our previous work where density functional learning via a local learning strategy was used in a variety of physical settings \cite{sammueller2023neural, sammueller2023whyNeural, sammueller2023neuralTutorial, delasheras2023perspective, zimmermann2024ml, sammueller2024hyperDFT, sammueller2024whyhyperDFT, sammueller2024pairmatching, sammueller2024attraction, buchanan2025attraction, kampa2024meta}, including the investigation of hard core correlation effects \cite{sammueller2023neural, sammueller2023whyNeural} and of interparticle attraction \cite{sammueller2024attraction, buchanan2025attraction} in describing gas-liquid phase separation in one-component systems.

We focus on a simple model binary mixture whose bulk phase behaviour has been investigated in many simulation and theoretical studies.
By contrast, the nature of fluid-fluid interfaces, the accompanying surface tensions, and wetting transitions have been addressed to a much smaller extent, with theories usually employing square-gradient or local density approximations \cite{telodagama1983one, telodagama1983two, tarazona1985interface, hadjiagapiou1985, schmid2001wetting, wilding2002, napari1999, rowlinsonwidombook}.
Setting the scene we note: i) it remains very difficult to progress beyond the naive mean-field incorporation of attraction within DFT, ii) predicting bulk phase behaviour on the basis of sophisticated integral equation theories remains challenging \cite{schoellpaschinger2003, antonevych2002}, and iii) carrying out accurate simulation work at interfaces in mixtures \cite{schmid2001wetting, wilding2002, martinezruiz2015, roy2024} is also challenging and requires systematic understanding of the bulk phase behaviour \cite{wilding1997, wilding1998, schmid2001wetting, wilding2002, koefinger2006epl, koefinger2006jcp, martinezruiz2015, roy2016, pathania2021}.

Here we adopt a new data-driven perspective to the physics of liquid-liquid phase separation, in which machine learning methods are deeply embedded and are inspired by DFT.
Specifically we consider the phase behaviour and interfacial structure of a \textit{symmetrical} model binary mixture characterized by Lennard-Jones (LJ) pair potentials $\phi_{ij}(r) = 4\epsilon_{ij}[(a_{ij}/r)^{12} - (a_{ij}/r)^6]$ acting between particles of species $i$ and $j$ where the species indices $i,j=1,2$.
We take $a_{11}=a_{22}=a_{12}=a$, i.e.\ identical diameters for each species, and consider weakened interparticle attraction between species 1 and 2, i.e.\ $\epsilon_{11}=\epsilon_{22}=\epsilon$ and $\epsilon_{12}<\epsilon$.
Each potential is truncated at $2.5a$.
Such a model has been investigated by several authors.
Our choice of parameters was motivated by those in the extensive simulation studies of Wilding et al.~\cite{wilding1997, schmid2001wetting, wilding2002} who investigated bulk phase behaviour and wetting at a particular solid substrate.
Their bulk phase diagrams point to the occurrence of what they and later authors term a $\lambda$~line, which is the line of critical transitions (upper consolute points) between a mixed (binary) fluid and a demixed (binary) fluid.
For a range of $\epsilon_{12}$ the $\lambda$~line meets the line of two-phase coexistence between the gas (vapour) and the liquid at a critical end point (CEP), occurring at the temperature $T_{\rm CEP}$.
Such behaviour is found in real fluid mixtures, type II in the important classification of \citet{vanKonynenburg1980}.
We choose $\epsilon_{12}=0.7\epsilon$ for which simulation studies \cite{wilding1998, schmid2001wetting, wilding2002} indicate that $T_{\rm CEP}$ lies well-below the gas-liquid critical temperature.
This information served as background for our training---see below---but did not directly influence our choice of parameter space.

\begin{figure}
\includegraphics[page=1,width=.49\textwidth]{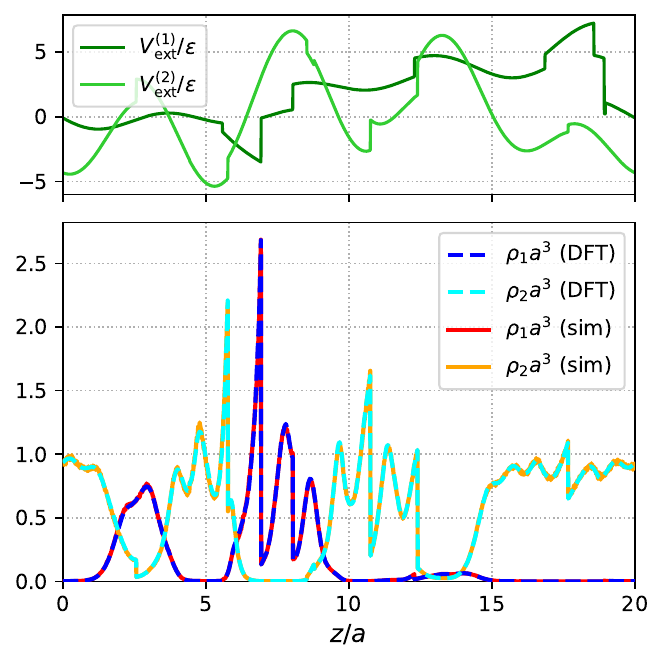}
\caption{
    To exemplify the generation of the training dataset, a specific realization of a randomized inhomogeneous environment is shown.
    This is characterized by the species-resolved external potentials $V_{\rmext}^{(1)}(z)/\epsilon$ and $V_{\rmext}^{(2)}(z)/\epsilon$ (top panel), chemical potentials $\mu_1/\epsilon =\mu_2/\epsilon=0.237599$ and temperature $k_BT/\epsilon=1.118673$.
    GCMC simulation results for the partial density profiles $\rho_1(z)$ and $\rho_2(z)$ are used in the local training of the species-resolved one-body direct correlation functional $c_1^{(i)}(z;[\rho_1,\rho_2],T)$, $i = 1, 2$; see text for details of the neural network.
    Whilst the training data consists of qualitatively similar profiles, the realization shown was not included in the training set thereby enabling a direct test of neural network predictions.
    The self-consistent solution of the Euler-Lagrange equations \eqref{EQselfConsistent} using the trained neural density functional yields results for the partial density profiles (labelled ``DFT'') that are identical on the scale of the plot to data generated by direct GCMC simulations (``sim'') for the specific external potentials; see bottom panel.
}
\label{FIGtraining}
\end{figure}

To apply the local learning scheme \cite{sammueller2023neural, sammueller2023whyNeural, sammueller2023neuralTutorial, sammueller2024pairmatching, sammueller2024attraction} to the LJ mixture, we use grand canonical Monte Carlo simulations (GCMC) to generate training data in the form of species-resolved (one-body) density profiles $\rho_1(z)$ and $\rho_2(z)$, which are inhomogeneous along a single coordinate $z$.
We consider planar geometry such that the system is translationally invariant in the two perpendicular $x$, $y$ directions.
The simulations are carried out for randomized values of temperature $T$ within the range $0.9 < k_B T / \epsilon < 2.0$ employing species-dependent, independently randomized forms of the external potentials $V_{\rmext}^{(1)}(z)$ and $V_{\rmext}^{(2)}(z)$ \cite{sammueller2023neural}.
Then the species-resolved Euler-Lagrange equations allow one to obtain the one-body direct correlation functions $c_1^{(1)}(z)$ and $c_1^{(2)}(z)$ according to
\begin{equation}
  c_1^{(i)}(z) = \ln \rho_i(z) + \beta V_\rmext^{(i)}(z)-\beta\mu_i,
  \label{EQc1training}
\end{equation}
where $i=1,2$, inverse temperature is $\beta=1/(k_BT)$, with Boltzmann constant $k_B$, and the thermal wavelengths are set to unity.
In all simulations, the species-dependent chemical potentials $\mu_i$ are set equal, such that $\mu_1=\mu_2=\mu$ in \eqr{EQc1training}, with randomized values of $\mu$ chosen uniformly within the range $-7 < \mu/\epsilon < 4$.
The species-resolved external potentials are constructed following the randomization process laid out in Ref.~\onlinecite{sammueller2023neural}.
Crucially, the randomization of $V_\mathrm{ext}^{(1)}(z)$ and $V_\mathrm{ext}^{(2)}(z)$ occurs independently in order to provide enough `contrast' between the two species, i.e.\ we deliberately avoid $V_\mathrm{ext}^{(1)}(z) = V_\mathrm{ext}^{(2)}(z)$.
This choice ensures that we probe sufficiently the relevant density inhomogeneities for the two species.

Having access to the pair of partial density profiles $\rho_1(z)$ and $\rho_2(z)$ together with the corresponding partial one-body direct correlation functions $c_1^{(1)}(z)$ and $c_1^{(2)}(z)$ \cite{evans1979, evans1992, evans2016} obtained via \eqr{EQc1training} allows training a neural network to represent the density functional relationship $c_1^{(i)}(z;[\rho_1,\rho_2],T)$; we indicate functional dependence by square brackets.
We represent this functional using the local learning scheme \cite{sammueller2023neural, sammueller2023whyNeural, sammueller2023neuralTutorial, sammueller2024pairmatching, sammueller2024attraction}, whereby a standard multilayer perceptron outputs both values of $c_1^{(i)}(z;[\rho_1,\rho_2])$, $i = 1, 2$, given as input the discretized partial density profiles $\rho_1(z'), \rho_2(z')$ within a window $|z' - z| < z_w$ around the position $z$ of interest.
We choose the spatial cutoff $z_w = 3.5 a$ following Ref.~\onlinecite{sammueller2024attraction}.
The dependence on temperature is captured by thermal training \cite{sammueller2024attraction, buchanan2025attraction, kampa2024meta} and by including the value of $T$ as an additional input node.
We note that a single neural network with two output nodes is used to yield values of $c_1^{(i)}(z)$ for both species $i=1,2$ simultaneously, which is in contrast to Ref.~\onlinecite{bui2024neuralrpm}.
While no symmetrization regarding the interchange of species is implemented directly in the neural network architecture, we use data augmentation during training and provide samples with flipped species indices as well as flipped on the $z$ axis to benefit from the underlying symmetry.
Fig.~\ref{FIGtraining} displays a typical choice of external potentials together with density profiles corresponding to a typical choice of thermodynamic parameters.

The trained neural one-body direct correlation functional $c_1^{(i)}(z;[\rho_1,\rho_2],T)$ encapsulates the effects of the interparticle interactions in the fluid mixture, as we will demonstrate.
In general, one can make ready and accurate predictions, for arbitrary external potentials, by self-consistent numerical solution of the species-resolved Euler-Lagrange equations:
\begin{equation}
  \rho_i(z) = \exp\big[ - \beta V_\rmext^{(i)}(z) + \beta\mu_i + c_1^{(i)}(z;[\rho_1,\rho_2],T) \big],
  \label{EQselfConsistent}
\end{equation}
which can be obtained formally from \eqr{EQc1training} by exponentiating and identifying the density functional dependence in $c_1^{(i)}(z;[\rho_1,\rho_2],T)$.
Equation \eqref{EQselfConsistent} is solved self-consistently via standard mixed Picard iteration.
Predictions for the partial density profiles from Eq.~\eqref{EQselfConsistent} with the neural density functional are highly accurate; see Fig.~\ref{FIGtraining} for a typical inhomogeneous situation, whereby Eq.~\eqref{EQselfConsistent} is solved using prescribed values of $\mu_1 = \mu_2 = \mu$ and $T$.

\begin{figure}
\includegraphics[page=1,width=.46\textwidth]{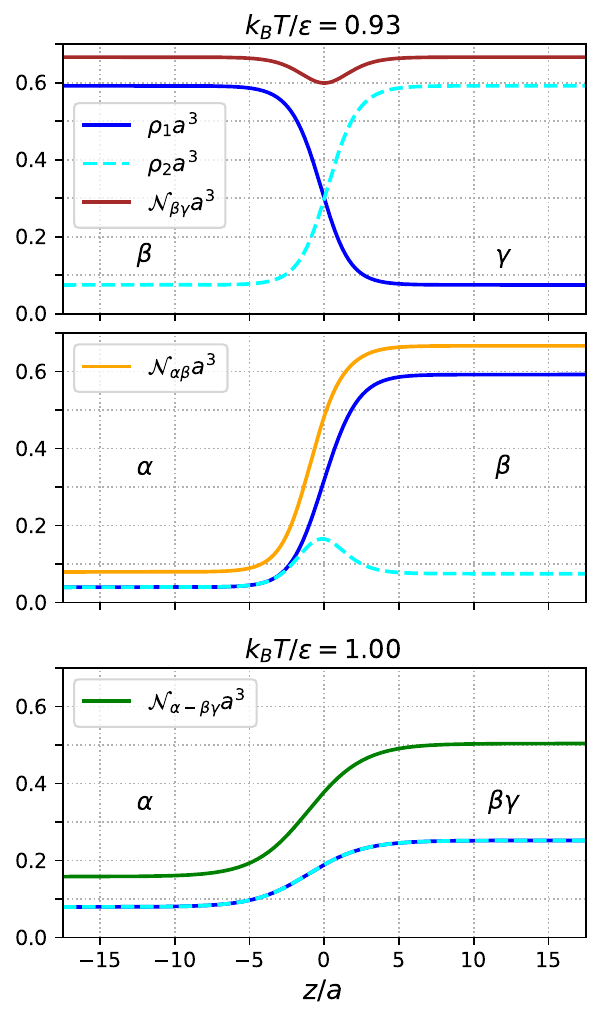} 
\caption{
    Density profiles at the different types of fluid-fluid interfaces predicted by the neural DFT: $\beta\gamma$ liquid-liquid (top panel) and $\alpha\beta$ gas-demixed liquid (second panel) at the scaled temperature $k_BT/\epsilon=0.93$.
    Shown are the scaled partial density profiles $\rho_1(z)a^3$ and $\rho_2(z)a^3$, as well as the scaled total density profile ${\cal N}(z)a^3=[\rho_1(z) + \rho_2(z)]a^3$.
    The $\alpha\gamma$ interface (not shown) is identical to the $\alpha\beta$ interface upon exchanging species 1 and 2.
    At the increased temperature $k_BT/\epsilon=1.0$ the system displays $\alpha$--$\beta\gamma$ coexistence (bottom panel) between $\alpha$ gas and mixed $\beta\gamma$ liquid with identical partial density profiles, $\rho_1(z) = \rho_2(z)$.
}
\label{FIGinterfaceStructure}
\end{figure}

We aim to apply the neural density functional to bulk phase coexistence and hence consider situations where both external potentials vanish, $V_\rmext^{(1)}(z) = V_\rmext^{(2)}(z) = 0$.
For stabilizing coexisting fluid states, we fix the mean value of one partial density via normalization in each iteration step.
Note that this determines implicitly the chemical potential $\mu_1 = \mu_2 = \mu$, which is equal for both species due to the employed symmetry.
Under appropriate initialization of the Picard iteration procedure to solve the Euler-Lagrange equation~\eqref{EQselfConsistent}, the theory predicts stable interfacial density profiles, where the partial density profiles cross over within an interfacial region between differing plateau (bulk) values.
We find that, depending on the statepoint chosen and the constraints employed within the iteration, all expected types of fluid-fluid coexistence emerge as solutions of \eqr{EQselfConsistent}.
We first display results for $k_BT/\epsilon=0.93$ (top and middle panels in Fig.~\ref{FIGinterfaceStructure}), where besides the partial density profiles $\rho_1(z)$ and $\rho_2(z)$, we also show the total density ${\cal N}(z)=\rho_1(z)+\rho_2(z)$.

\begin{figure}
\includegraphics[page=1,width=.49\textwidth]{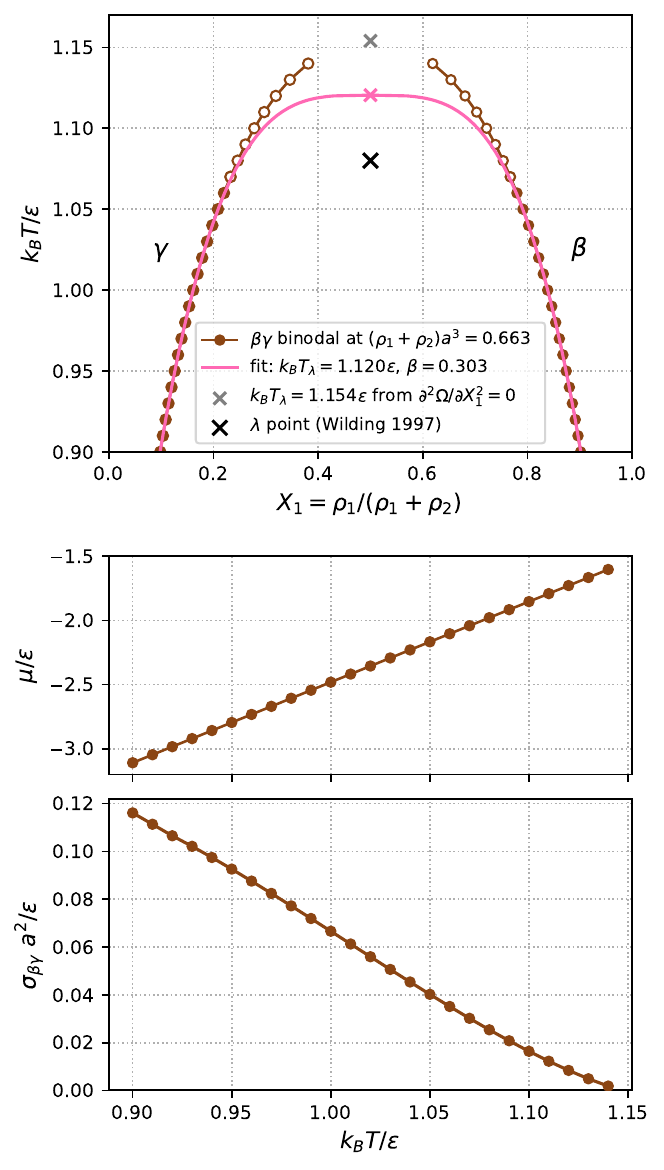}
\caption{
    Top panel: neural density functional results for the bulk $\beta\gamma$ liquid-liquid binodal at fixed total bulk density $(\rho_1+\rho_2)a^3=0.663$ shown as a function of bulk composition $X_1=\rho_1/(\rho_1+\rho_2)$ and scaled temperature $k_BT/\epsilon$.
    The results (circles) are obtained from the plateau values of equilibrium interfacial density profiles (see the top panel in Fig.~\ref{FIGinterfaceStructure}).
    The fit according to \eqr{EQfitFunctionShortPaper} is obtained using only data below and including cutoff temperature $k_BT/\epsilon=1.06$ (brown circles), with the resulting numerical values for $k_BT_\lambda/\epsilon$  and exponent $\beta$ given in the legend.
    For comparison, the black cross denotes the $\lambda$~point obtained from simulation by \citet{wilding1997}.
    Middle panel: variation of the scaled chemical potential $\mu / \varepsilon = \mu_1 / \varepsilon = \mu_2 / \varepsilon$ with respect to temperature along the specified path of constant total density.
    Bottom panel: neural density functional results for the scaled tension $\sigma_{\beta\gamma} a^2 / \epsilon$ of the $\beta\gamma$ interface obtained from functional line integration \eqref{EQfunctionalIntegral}.
}
\label{FIGliqliqBinodal}
\end{figure}

At this (low) temperature we find three distinct interfaces: (i) $\alpha\beta$, where $\alpha$ is a gas and $\beta$ a liquid with majority component 1, (ii) $\alpha\gamma$, where $\gamma$ is a liquid with majority component 2, and (iii) $\beta\gamma$, the liquid-liquid interface.
Note that the $\alpha\beta$ interface displays an adsorption maximum for the profile of species 2 and that the $\alpha\gamma$ interface (not shown) displays, respecting symmetry, an equivalent maximum for species 1.
By contrast the density profiles of both species are monotonic at the $\beta\gamma$ interface.
Such behaviour was suggested in early DFT calculations \cite{telodagama1983one}, albeit for a weakly asymmetrical mixture.

The existence of three ($\alpha\beta$, $\alpha\gamma$, and $\beta\gamma$) interfaces for temperature $k_B T / \varepsilon = 0.93$, see Fig.~\ref{FIGinterfaceStructure}, implies this particular statepoint lies on the three (fluid) phase line of our model.
Indeed reading off the plateau values of the density profiles in Fig.~\ref{FIGinterfaceStructure} yields values for the bulk densities of each species in the coexisting phases.
On raising the temperature, we reach a point where the density profiles at the $\beta\gamma$ interface become identical and flat and we associate this with the critical end point $T_{\rm CEP}$.
For higher temperatures, $T>T_{\rm CEP}$, there is no longer demixing between the two liquid phases and gas ($\alpha$) coexists with a mixed ($\beta\gamma$) liquid phase.
There is a single ${\alpha-\beta\gamma}$ interface; the bottom panel of Fig.~\ref{FIGinterfaceStructure} with $k_BT/\epsilon = 1.00$ provides an example.
We return to the bulk phase diagram later but first turn attention to the liquid-liquid ($\beta\gamma$) interfaces and coexistence that we determine at total densities larger than those in Fig.~\ref{FIGinterfaceStructure}.
Specifically, we choose to fix the total bulk density $\rho_1+\rho_2$ and increase $T$.
The symmetry of our model LJ liquid mixture dictates that the two demixed liquid phases must exhibit a symmetrical Ising-like coexistence curve when expressed in terms of a composition variable $X_1=\rho_1/(\rho_1+\rho_2)$, with the upper critical point at $X_1=1/2$.

\begin{figure}
\includegraphics[page=1,width=.44\textwidth]{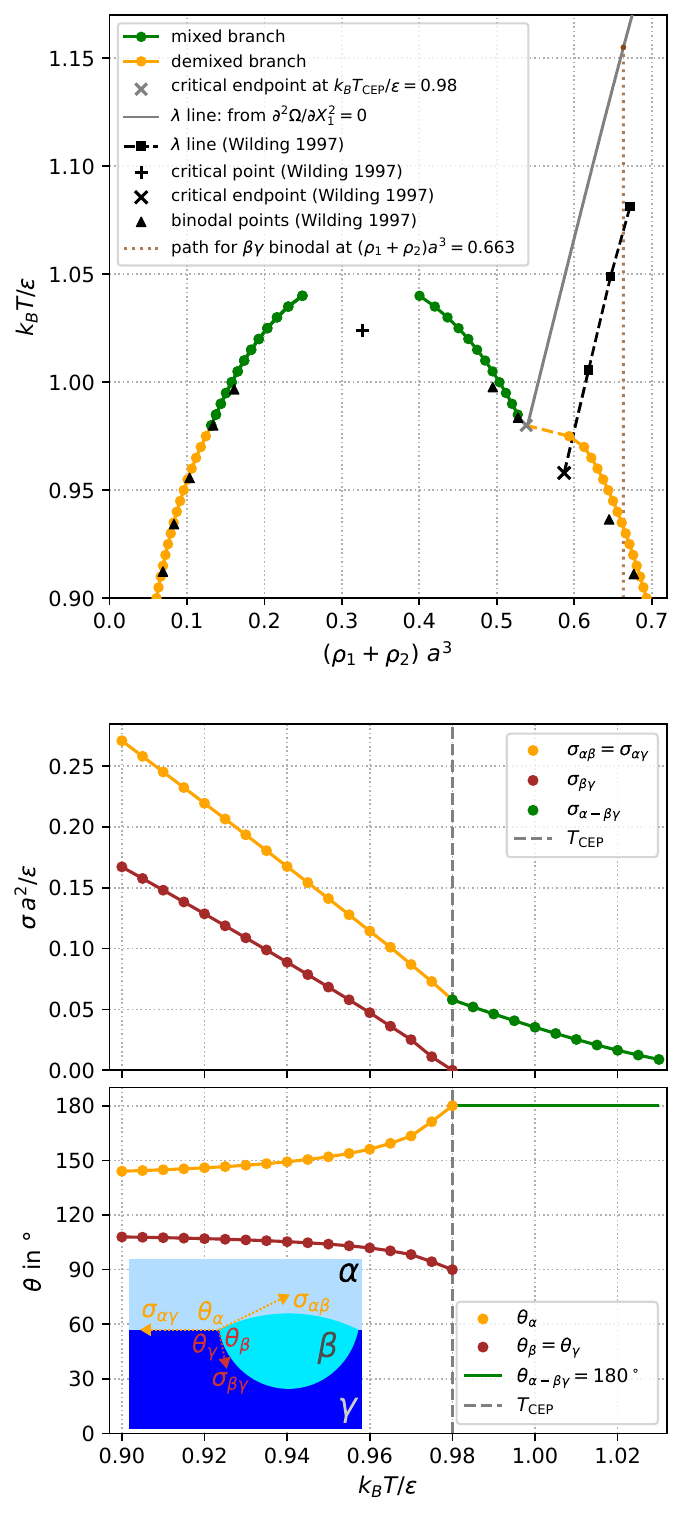}
\caption{
    Top panel: bulk fluid phase diagram of the symmetrical binary Lennard-Jones system as a function of scaled total density $(\rho_1+\rho_2)a^3$ and scaled temperature $k_BT/\epsilon$ together with simulation data of \citet{wilding1997}.
    Binodals and the $\lambda$~line are plotted.
    The vertical dotted red line indicates the path of constant total density for which the liquid-liquid binodal is shown in Fig.~\ref{FIGliqliqBinodal}.
    Middle panel: scaled interfacial tensions between gas and demixed liquid, $\sigma_{\alpha\beta}a^2/\epsilon = \sigma_{\alpha\gamma}a^2/\epsilon$ and between liquid and liquid, $\sigma_{\beta\gamma}a^2/\epsilon$, for $T< T_{\rm CEP}$ and between the gas and the mixed liquid, $\sigma_{\alpha-\beta\gamma}a^2/\epsilon$ (green dots), for $T>T_{\rm CEP}$.
    The dashed vertical line indicates $T_{\rm CEP}$.
    Bottom panel: contact angles $\theta_\alpha$ and $\theta_\beta=\theta_\gamma$ on the triple phase line, obtained from \eqr{EQcosThetaBeta}.
    The sketch shows the lens of (demixed) liquid $\beta$ together with the contact angles and tensions.
}
\label{FIGcontactangle}
\end{figure}

In Fig.~\ref{FIGliqliqBinodal} (top panel), we plot the coexistence values of $X_1$ as a function of temperature, while keeping the total bulk density $\rho_1 + \rho_2 = 0.663 a^{-3}$ fixed.
Recall that $\mu_1 = \mu_2 = \mu$ holds due to symmetry, but that the value of $\mu$ changes, as expected, along the path of constant total density upon varying $T$, as shown in Fig.~\ref{FIGliqliqBinodal} (middle panel).
The two liquid branches meet at an upper critical temperature $T_\lambda$ at $X_1=1/2$.
As an alternative to tracing out the liquid-liquid binodal and determining the merging point of both branches, $T_{\lambda}$ can also be identified as the point of vanishing second derivative of the grand potential $\Omega$ with respect to composition $X_1$; see grey cross in Fig.~\ref{FIGliqliqBinodal}.
Automatic differentiation of the neural functional enables this calculation to be performed efficiently \cite{sammueller2023neural}, which allows mapping out the line of $\lambda$~points in the phase diagram at low computational cost.
We return to this in Fig.~\ref{FIGcontactangle}, but proceed first with what should be a more accurate estimation of the value $T_\lambda$, taking into account the subtleties that arise when evaluating the neural functional close to critical points \cite{sammueller2024attraction}.

Specifically we adapt the fitting method used for the gas-liquid binodal in the pure LJ system \cite{sammueller2024attraction} and use the empirical Ising/lattice gas scaling form for this particular liquid-liquid phase separation:
\begin{equation}
  X_1 = \frac{1}{2} \, \pm \, b |T^\ast - T_\lambda^\ast|^\beta,
  \label{EQfitFunctionShortPaper}
\end{equation}
where $T^\ast = k_BT/\epsilon$ is scaled temperature, $b$ is a constant, the critical concentration is $1/2$ per symmetry, and the exponent $\beta$ should neither be confused with inverse temperature nor with the liquid phase in which species 1 is the majority component.
Because of the Ising-like symmetry there is no need to include a linear contribution in \eqr{EQfitFunctionShortPaper}.
The three-dimensional Ising value is $\beta = 0.326 30(22)$ \cite{ferrenberg2018}, but here the exponent $\beta$ is treated as a free fit parameter.
Following Ref.~\onlinecite{sammueller2024attraction} we exclude data very close to the critical $\lambda$~point, using only coexisting densities for $k_B T / \varepsilon \leq 1.06$, and find from the fitting blue$T_\lambda^\ast = 1.120$ and $\beta=0.303$ for this particular total density, see Fig.~\ref{FIGliqliqBinodal}.
$T_{\lambda}$ obtained from the fitting procedure is about $3\%$ lower than the value obtained from direct evaluation of the neural functional by tracing the $\beta \gamma$ binodal, i.e.\ calculating the merging of coexisting density profiles or, equivalently, from the vanishing of the second-derivative of the grand potential.

Our framework allows us to access interfacial tensions via expressing the excess free energy difference $\Delta F_\rmexc$ as a functional line integral,
\begin{equation}
  \frac{\beta\Delta F_\rmexc}{A} = - \int dz \sum_{i=1,2} \Delta\rho_i(z) \int_0^1 ds \ c_1^{(i)}(z;[\rho_{1,s},\rho_{2,s}],T),
  \label{EQfunctionalIntegral}
\end{equation}
where $A$ is the lateral area of the system and the density profiles $\rho_{1,s}(z)$ and $\rho_{2,s}(z)$ used as functional arguments are taken to be superpositions $\rho_{i,s}(z)= s \rho_i(z) + (1-s) \rho_i$, where $\rho_i$ is the bulk coexistence density for species $i=1,2$ and the difference between start and end densities is $\Delta\rho_i(z) = \rho_{i,1}(z)-\rho_{i,0}(z)=\rho_i(z)-\rho_i$.
Then the surface tension is obtained from the excess grand potential $\sigma = (\Omega + pV)/A$, where $p$ is the pressure at coexistence.
The results, shown in Fig.~\ref{FIGliqliqBinodal} (bottom panel), indicate as expected that $\sigma_{\beta\gamma}$ decreases monotonically with temperature and that it vanishes as $T\to T_\lambda$.

We return now to the full phase diagram obtained from considering density profiles such as those in Fig.~\ref{FIGinterfaceStructure}.
Analysis of the plateau values of all stable fluid interfaces allows us to construct the entire fluid phase diagram, shown in the top panel of Fig.~\ref{FIGcontactangle}, and plotted as a function of the total density $\rho_1+\rho_2$ and temperature $T$.
We have checked that for $T > T_{\rm CEP}$ the results obtained for coexisting densities are numerically consistent with those obtained via the Maxwell construction based on bulk functional integration \cite{sammueller2023neural}.
The representation in Fig.~\ref{FIGcontactangle} makes apparent the density jump at gas-liquid coexistence.
Recall that above the CEP the liquid is in a mixed state; both species have equal concentration.
Below the CEP the liquid is demixed, as discussed above, and the concentration jump (see Fig.~\ref{FIGliqliqBinodal}) is `collapsed' in the representation of Fig.~\ref{FIGcontactangle}.
As described above, the $\lambda$~line is calculated from the locus of vanishing second derivative of the grand potential $\Omega$ with respect to composition $X_1$, readily facilitated via automatic differentiation of the neural functional \cite{sammueller2023neural}.
Whilst our results for the gas-liquid binodal agree closely with the independent simulation data of Wilding \cite{wilding1997}, our $\lambda$~line displays a small offset, typically about $10\%$ in density, with respect to the corresponding simulation results.

We obtain the surface tensions for gas-liquid coexistence again via \eqr{EQfunctionalIntegral}, adapted to the respective bulk coexistence conditions.
Figure~\ref{FIGcontactangle} (middle panel) displays results obtained below $T_{\rm CEP}$ for the surface tensions of the gas-liquid interfaces, $\sigma_{\alpha\beta}=\sigma_{\alpha\gamma}$, and for the liquid-liquid interface, $\sigma_{\beta\gamma}$.
The latter vanishes at $T_{\rm CEP}$, cf.\ Fig.~\ref{FIGliqliqBinodal}.
For $T>T_{\rm CEP}$ the surface tension refers to the gas-liquid interface, $\sigma_{\alpha-\beta\gamma}$, where the liquid is now in a mixed state.

At triple phase coexistence below $T_{\rm CEP}$, three fluid phases can meet in stable mechanical contact and form a lens of, say, the liquid $\beta$ phase.
The corresponding contact angles $\theta_\alpha$, $\theta_\beta$, and $\theta_\gamma$ characterize the (macroscopic) shape of the lens and satisfy $\theta_\alpha + \theta_\beta + \theta_\gamma = 2\pi$, see e.g.\ Ref.~\onlinecite{rowlinsonwidombook}.
Generally, the contact angles are determined by the three surface tensions via the well-known Neumann triangle construction \cite{rowlinsonwidombook}.
The symmetry of our LJ mixture dictates that the surface tensions $\sigma_{\alpha\beta}=\sigma_{\alpha\gamma}$, so that the triangle is isosceles and $\theta_\beta = \theta_\gamma$.
It follows that 
\begin{equation}
  \cos\theta_\beta = -\frac{\sigma_{\beta\gamma}}{2\sigma_{\alpha\beta}}.
  \label{EQcosThetaBeta}
\end{equation}
Only the ratio of the respective interfacial tensions is relevant.
The form of the liquid lens, as determined by the three tensions, together with results for the contact angles as a function of temperature, are shown in the bottom panel of Fig.~\ref{FIGcontactangle}.
Approaching $T_{\rm CEP}$ from below we find as expected: $\theta_\beta=\theta_\gamma \to \pi/2$ and $\theta_\alpha \to \pi$.

For our \emph{symmetrical} model there can be no transition to complete wetting of the $\alpha\gamma$ interface by liquid $\beta$ for $T<T_{\rm CEP}$: the right-hand side of Eq.~\eqref{EQcosThetaBeta} can only vanish at $T_{\rm CEP}$.
This implies partial wetting, i.e.\ a finite lens, rather than the intervention of a thick film of phase $\beta$, pertains for all $T< T_{\rm CEP}$---an observation pertinent for the general physics of wetting transitions at fluid-fluid interfaces.
The latter topic continues to attract considerable attention, see e.g.\ \citet{parry2024}, \citet{indekeu2022}, and references therein.
Remarks: i) the consequences for wetting imposed by symmetry of the interparticle potentials were already implicit in Sec.~8.3 of Ref.~\onlinecite{rowlinsonwidombook}, ii) breaking the symmetry can lead to first order \cite{telodagama1983two} or critical \cite{parry2024} wetting transitions below $T_{\rm CEP}$, and iii) determining accurately the behaviour of the surface tensions close to $T_{\rm CEP}$ is very difficult within the neural functional framework \cite{sammueller2024attraction}; similar considerations apply to direct simulation studies.

In conclusion, we have investigated the bulk fluid phase behaviour and associated interfacial phenomena in a symmetrical LJ mixture with a specific choice of weakened cross attraction.
Whilst imposing symmmetry serves conveniently to reduce the parameter space, the model already incorporates a wealth of physics that is challenging to describe within a machine learning context: important is the lack of \emph{a priori} symmetry breaking.
A single neural network was used to represent the pair of species-resolved one-body direct correlation functionals simultaneously.
This strategy is consistent with the formal generation of $c_1^{(i)}(\rv;[\rho_1,\rho_2],T)$ via functional differentiation from a single, underlying, excess free energy functional.

Training the neural functional is based on adapting both the local learning \cite{sammueller2023neural} and thermal training \cite{sammueller2024attraction} schemes that were previously developed for single-component systems. Using these methods directly for the present binary mixture allows one to learn solely on the basis of the available simulation data.
We have found that small artifacts can arise thereby in the predictions of the precise shape and location of the gas-liquid  binodal when using the above schemes without modification.
These artifacts are prevented by including $L_2$ regularization\footnote{Using $L_2$ regularization during training is a standard practice in machine learning which penalizes large values in the neural network parameters. This supports the generation of less complex (`smooth') models, which is particularly suitable for the functional mapping $c_1^{(i)}(\vec{r}; [\rho_1, \rho_2], T)$.} in the training protocol, which we have hence adopted for the neural functional that was used to generate all numerical results shown in this work.
The two different neural models trained with and without $L_2$ regularization, together with a comparison of the corresponding numerical predictions, are available in Ref.~\onlinecite{Zenodo}.
In future work, it would be interesting to investigate physics-informed regularization, e.g.\ by incorporating exact statistical mechanical sum rules\cite{hermann2021} to provide additional constraints that are to be optimized during training.

Working with the trained neural density functional is computationally more efficient than carrying out direct simulations to determine the (bulk) phase diagram.
Moreover, we demonstrated that the methodology provides systematic access to subtle interfacial properties.
It would be very interesting for simulation experts to test the predictions for surface tensions and contact angles in Fig.~\ref{FIGcontactangle}.
Determining these remains a significant challenge for direct simulation work.
The benefits of the neural method far outweigh the very moderate computational overhead for creating the training data \cite{sammueller2025mu}, which involves standard, and cheap, simulations of inhomogeneous one-body density profiles of sufficient variability, straightforward to ensure in practice using appropriate random external potentials.
Predictions made with the neural functional are numerically robust and agree quantitatively with reference data for statepoints away from criticality.
Overcoming limitations close to critical points, where correlation effects on large length scales related to critical fluctuations become relevant, remain to be properly incorporated.

Functional differentiation provides access to the partial two-body direct correlation functions \cite{sammueller2023neural, sammueller2024attraction} and, hence via the Ornstein-Zernike route, to partial bulk and interfacial structure factors which provide further insight into the physics of liquid-liquid phase separation and the nature of the line of upper consolute points ($\lambda$~line).
For example, at points on this line all three bulk structure factors $S_{ij}(k)$ diverge at wave number $k=0$ but the isothermal compressibility does not diverge.
Results will be presented elsewhere.

It is compelling to apply our methodology to broader types of fluid mixtures.
For example, simply changing the cross parameter gives rise to a tricritical point for sufficiently large $\varepsilon_{12}$, and intricate wetting behaviour is expected to be found for the asymmetric case, $\varepsilon_{11} \neq \varepsilon_{22}$.
These scenariors can all be studied with neural density functionals.
We expect our machine learning technique to fare very well with the increased parameter space \cite{kampa2024meta}, provided that sufficient training data can be generated.
A chemical engineer might inquire how well we can address realistic (chemical) substances.
The applicability of our method clearly hinges upon the availability of reference data and their effective incorporation into the framework of DFT \cite{klink2014,rehner2023,bursik2025}.
The present paper focuses on the underlying physics of the approach.

The neural density functional is available online, and we recall its universal applicability in planar inhomogeneous situations, cf.\ Fig.~\ref{FIGtraining}.

\begin{acknowledgments}
We are grateful to A.~O.~Parry and N.~B.~Wilding for valuable conversations and for providing references.
Some of the calculations were performed using the emil-cluster of the Bayreuth Centre for High Performance Computing funded by the DFG (Deutsche Forschungsgemeinschaft) under Project No.~422127126.
This work is supported by the DFG (Deutsche Forschungsgemeinschaft) under Project No.~551294732.
\end{acknowledgments}

\section*{Data Availability Statement}

The data that support the findings of this study are openly available \cite{Zenodo}.

\bibliography{bibliography}

\begin{thebibliography}{75}%
\makeatletter
\providecommand \@ifxundefined [1]{%
 \@ifx{#1\undefined}
}%
\providecommand \@ifnum [1]{%
 \ifnum #1\expandafter \@firstoftwo
 \else \expandafter \@secondoftwo
 \fi
}%
\providecommand \@ifx [1]{%
 \ifx #1\expandafter \@firstoftwo
 \else \expandafter \@secondoftwo
 \fi
}%
\providecommand \natexlab [1]{#1}%
\providecommand \enquote  [1]{``#1''}%
\providecommand \bibnamefont  [1]{#1}%
\providecommand \bibfnamefont [1]{#1}%
\providecommand \citenamefont [1]{#1}%
\providecommand \href@noop [0]{\@secondoftwo}%
\providecommand \href [0]{\begingroup \@sanitize@url \@href}%
\providecommand \@href[1]{\@@startlink{#1}\@@href}%
\providecommand \@@href[1]{\endgroup#1\@@endlink}%
\providecommand \@sanitize@url [0]{\catcode `\\12\catcode `\$12\catcode
  `\&12\catcode `\#12\catcode `\^12\catcode `\_12\catcode `\%12\relax}%
\providecommand \@@startlink[1]{}%
\providecommand \@@endlink[0]{}%
\providecommand \url  [0]{\begingroup\@sanitize@url \@url }%
\providecommand \@url [1]{\endgroup\@href {#1}{\urlprefix }}%
\providecommand \urlprefix  [0]{URL }%
\providecommand \Eprint [0]{\href }%
\providecommand \doibase [0]{http://dx.doi.org/}%
\providecommand \selectlanguage [0]{\@gobble}%
\providecommand \bibinfo  [0]{\@secondoftwo}%
\providecommand \bibfield  [0]{\@secondoftwo}%
\providecommand \translation [1]{[#1]}%
\providecommand \BibitemOpen [0]{}%
\providecommand \bibitemStop [0]{}%
\providecommand \bibitemNoStop [0]{.\EOS\space}%
\providecommand \EOS [0]{\spacefactor3000\relax}%
\providecommand \BibitemShut  [1]{\csname bibitem#1\endcsname}%
\let\auto@bib@innerbib\@empty
\bibitem [{\citenamefont {Frenkel}\ and\ \citenamefont
  {Smit}(2023)}]{frenkel2023book}%
  \BibitemOpen
  \bibfield  {author} {\bibinfo {author} {\bibfnamefont {D.}~\bibnamefont
  {Frenkel}}\ and\ \bibinfo {author} {\bibfnamefont {B.}~\bibnamefont {Smit}},\
  }\href@noop {} {\emph {\bibinfo {title} {Understanding Molecular Simulation:
  From Algorithms to Applications}}},\ \bibinfo {edition} {3rd}\ ed.\ (\bibinfo
   {publisher} {Academic Press},\ \bibinfo {address} {Amsterdam},\ \bibinfo
  {year} {2023})\BibitemShut {NoStop}%
\bibitem [{\citenamefont {van Konynenburg}\ and\ \citenamefont
  {Scott}(1980)}]{vanKonynenburg1980}%
  \BibitemOpen
  \bibfield  {author} {\bibinfo {author} {\bibfnamefont {P.~H.}\ \bibnamefont
  {van Konynenburg}}\ and\ \bibinfo {author} {\bibfnamefont {R.~L.}\
  \bibnamefont {Scott}},\ }\bibfield  {title} {\enquote {\bibinfo {title}
  {Critical lines and phase equilibria in binary van der {{Waals}} mixtures},}\
  }\href {\doibase 10.1098/rsta.1980.0266} {\bibfield  {journal} {\bibinfo
  {journal} {Philos. Trans. R. Soc. Ser. A}\ }\textbf {\bibinfo {volume}
  {298}},\ \bibinfo {pages} {495--540} (\bibinfo {year} {1980})}\BibitemShut
  {NoStop}%
\bibitem [{\citenamefont {Sciortino}\ \emph {et~al.}(2025)\citenamefont
  {Sciortino}, \citenamefont {Zhai}, \citenamefont {Bore},\ and\ \citenamefont
  {Paesani}}]{sciortino2025}%
  \BibitemOpen
  \bibfield  {author} {\bibinfo {author} {\bibfnamefont {F.}~\bibnamefont
  {Sciortino}}, \bibinfo {author} {\bibfnamefont {Y.}~\bibnamefont {Zhai}},
  \bibinfo {author} {\bibfnamefont {S.~L.}\ \bibnamefont {Bore}}, \ and\
  \bibinfo {author} {\bibfnamefont {F.}~\bibnamefont {Paesani}},\ }\bibfield
  {title} {\enquote {\bibinfo {title} {Constraints on the location of the
  liquid--liquid critical point in water},}\ }\href {\doibase
  10.1038/s41567-024-02761-0} {\bibfield  {journal} {\bibinfo  {journal} {Nat.
  Phys.}\ }\textbf {\bibinfo {volume} {21}},\ \bibinfo {pages} {480--485}
  (\bibinfo {year} {2025})}\BibitemShut {NoStop}%
\bibitem [{\citenamefont {Archer}\ \emph {et~al.}(2024)\citenamefont {Archer},
  \citenamefont {Goddard}, \citenamefont {Sibley}, \citenamefont {Rawlings},
  \citenamefont {Broadhurst}, \citenamefont {Ouali},\ and\ \citenamefont
  {Fairhurst}}]{archer2024ouzo}%
  \BibitemOpen
  \bibfield  {author} {\bibinfo {author} {\bibfnamefont {A.~J.}\ \bibnamefont
  {Archer}}, \bibinfo {author} {\bibfnamefont {B.~D.}\ \bibnamefont {Goddard}},
  \bibinfo {author} {\bibfnamefont {D.~N.}\ \bibnamefont {Sibley}}, \bibinfo
  {author} {\bibfnamefont {J.~T.}\ \bibnamefont {Rawlings}}, \bibinfo {author}
  {\bibfnamefont {R.}~\bibnamefont {Broadhurst}}, \bibinfo {author}
  {\bibfnamefont {F.~F.}\ \bibnamefont {Ouali}}, \ and\ \bibinfo {author}
  {\bibfnamefont {D.~J.}\ \bibnamefont {Fairhurst}},\ }\bibfield  {title}
  {\enquote {\bibinfo {title} {Experimental and theoretical bulk phase diagram
  and interfacial tension of ouzo},}\ }\href {\doibase 10.1039/D4SM00332B}
  {\bibfield  {journal} {\bibinfo  {journal} {Soft Matter}\ }\textbf {\bibinfo
  {volume} {20}},\ \bibinfo {pages} {5889--5903} (\bibinfo {year}
  {2024})}\BibitemShut {NoStop}%
\bibitem [{\citenamefont {Sibley}\ \emph {et~al.}(2025)\citenamefont {Sibley},
  \citenamefont {Goddard}, \citenamefont {Ouali}, \citenamefont {Fairhurst},\
  and\ \citenamefont {Archer}}]{sibley2025}%
  \BibitemOpen
  \bibfield  {author} {\bibinfo {author} {\bibfnamefont {D.~N.}\ \bibnamefont
  {Sibley}}, \bibinfo {author} {\bibfnamefont {B.~D.}\ \bibnamefont {Goddard}},
  \bibinfo {author} {\bibfnamefont {F.~F.}\ \bibnamefont {Ouali}}, \bibinfo
  {author} {\bibfnamefont {D.~J.}\ \bibnamefont {Fairhurst}}, \ and\ \bibinfo
  {author} {\bibfnamefont {A.~J.}\ \bibnamefont {Archer}},\ }\bibfield  {title}
  {\enquote {\bibinfo {title} {Coexisting multiphase and interfacial behavior
  of ouzo},}\ }\href {\doibase 10.1063/5.0253815} {\bibfield  {journal}
  {\bibinfo  {journal} {Phys. Fluids}\ }\textbf {\bibinfo {volume} {37}},\
  \bibinfo {pages} {042118} (\bibinfo {year} {2025})}\BibitemShut {NoStop}%
\bibitem [{\citenamefont {Hyman}, \citenamefont {Weber},\ and\ \citenamefont
  {J{\"u}licher}(2014)}]{hyman2014}%
  \BibitemOpen
  \bibfield  {author} {\bibinfo {author} {\bibfnamefont {A.~A.}\ \bibnamefont
  {Hyman}}, \bibinfo {author} {\bibfnamefont {C.~A.}\ \bibnamefont {Weber}}, \
  and\ \bibinfo {author} {\bibfnamefont {F.}~\bibnamefont {J{\"u}licher}},\
  }\bibfield  {title} {\enquote {\bibinfo {title} {Liquid-{{Liquid Phase
  Separation}} in {{Biology}}},}\ }\href {\doibase
  10.1146/annurev-cellbio-100913-013325} {\bibfield  {journal} {\bibinfo
  {journal} {Annu. Rev. Cell Dev. Biol.}\ }\textbf {\bibinfo {volume} {30}},\
  \bibinfo {pages} {39--58} (\bibinfo {year} {2014})}\BibitemShut {NoStop}%
\bibitem [{\citenamefont {Arnold}\ \emph {et~al.}(2024)\citenamefont {Arnold},
  \citenamefont {Sch{\"a}fer}, \citenamefont {Edelman},\ and\ \citenamefont
  {Bruder}}]{arnold2024}%
  \BibitemOpen
  \bibfield  {author} {\bibinfo {author} {\bibfnamefont {J.}~\bibnamefont
  {Arnold}}, \bibinfo {author} {\bibfnamefont {F.}~\bibnamefont {Sch{\"a}fer}},
  \bibinfo {author} {\bibfnamefont {A.}~\bibnamefont {Edelman}}, \ and\
  \bibinfo {author} {\bibfnamefont {C.}~\bibnamefont {Bruder}},\ }\bibfield
  {title} {\enquote {\bibinfo {title} {Mapping out phase diagrams with
  generative classifiers},}\ }\href {\doibase 10.1103/PhysRevLett.132.207301}
  {\bibfield  {journal} {\bibinfo  {journal} {Phys. Rev. Lett.}\ }\textbf
  {\bibinfo {volume} {132}},\ \bibinfo {pages} {207301} (\bibinfo {year}
  {2024})}\BibitemShut {NoStop}%
\bibitem [{\citenamefont {Bedolla}, \citenamefont {Padierna},\ and\
  \citenamefont {{Casta{\~n}eda-Priego}}(2021)}]{bedolla2021}%
  \BibitemOpen
  \bibfield  {author} {\bibinfo {author} {\bibfnamefont {E.}~\bibnamefont
  {Bedolla}}, \bibinfo {author} {\bibfnamefont {L.~C.}\ \bibnamefont
  {Padierna}}, \ and\ \bibinfo {author} {\bibfnamefont {R.}~\bibnamefont
  {{Casta{\~n}eda-Priego}}},\ }\bibfield  {title} {\enquote {\bibinfo {title}
  {Machine learning for condensed matter physics},}\ }\href {\doibase
  10.1088/1361-648X/abb895} {\bibfield  {journal} {\bibinfo  {journal} {J.
  Phys.: Condens. Matter}\ }\textbf {\bibinfo {volume} {33}},\ \bibinfo {pages}
  {053001} (\bibinfo {year} {2021})}\BibitemShut {NoStop}%
\bibitem [{\citenamefont {Chertenkov}, \citenamefont {Burovski},\ and\
  \citenamefont {Shchur}(2023)}]{chertenkov2023}%
  \BibitemOpen
  \bibfield  {author} {\bibinfo {author} {\bibfnamefont {V.}~\bibnamefont
  {Chertenkov}}, \bibinfo {author} {\bibfnamefont {E.}~\bibnamefont
  {Burovski}}, \ and\ \bibinfo {author} {\bibfnamefont {L.}~\bibnamefont
  {Shchur}},\ }\bibfield  {title} {\enquote {\bibinfo {title} {Finite-size
  analysis in neural network classification of critical phenomena},}\ }\href
  {\doibase 10.1103/PhysRevE.108.L032102} {\bibfield  {journal} {\bibinfo
  {journal} {Phys. Rev. E}\ }\textbf {\bibinfo {volume} {108}},\ \bibinfo
  {pages} {L032102} (\bibinfo {year} {2023})}\BibitemShut {NoStop}%
\bibitem [{\citenamefont {Dijkstra}\ and\ \citenamefont
  {Luijten}(2021)}]{dijkstra2021ml}%
  \BibitemOpen
  \bibfield  {author} {\bibinfo {author} {\bibfnamefont {M.}~\bibnamefont
  {Dijkstra}}\ and\ \bibinfo {author} {\bibfnamefont {E.}~\bibnamefont
  {Luijten}},\ }\bibfield  {title} {\enquote {\bibinfo {title} {From predictive
  modelling to machine learning and reverse engineering of colloidal
  self-assembly},}\ }\href {\doibase 10.1038/s41563-021-01014-2} {\bibfield
  {journal} {\bibinfo  {journal} {Nat. Mater.}\ }\textbf {\bibinfo {volume}
  {20}},\ \bibinfo {pages} {762--773} (\bibinfo {year} {2021})}\BibitemShut
  {NoStop}%
\bibitem [{\citenamefont {Coli}\ \emph {et~al.}(2022)\citenamefont {Coli},
  \citenamefont {Boattini}, \citenamefont {Filion},\ and\ \citenamefont
  {Dijkstra}}]{coli2022scienceAdvances}%
  \BibitemOpen
  \bibfield  {author} {\bibinfo {author} {\bibfnamefont {G.~M.}\ \bibnamefont
  {Coli}}, \bibinfo {author} {\bibfnamefont {E.}~\bibnamefont {Boattini}},
  \bibinfo {author} {\bibfnamefont {L.}~\bibnamefont {Filion}}, \ and\ \bibinfo
  {author} {\bibfnamefont {M.}~\bibnamefont {Dijkstra}},\ }\bibfield  {title}
  {\enquote {\bibinfo {title} {Inverse design of soft materials via a deep
  learning--based evolutionary strategy},}\ }\href {\doibase
  10.1126/sciadv.abj6731} {\bibfield  {journal} {\bibinfo  {journal} {Sci.
  Adv.}\ }\textbf {\bibinfo {volume} {8}},\ \bibinfo {pages} {eabj6731}
  (\bibinfo {year} {2022})}\BibitemShut {NoStop}%
\bibitem [{\citenamefont {Liu}\ \emph {et~al.}(2024)\citenamefont {Liu},
  \citenamefont {Zhang}, \citenamefont {Ru}, \citenamefont {Gao}, \citenamefont
  {Moore},\ and\ \citenamefont {Yan}}]{liu2024}%
  \BibitemOpen
  \bibfield  {author} {\bibinfo {author} {\bibfnamefont {Z.}~\bibnamefont
  {Liu}}, \bibinfo {author} {\bibfnamefont {X.}~\bibnamefont {Zhang}}, \bibinfo
  {author} {\bibfnamefont {X.}~\bibnamefont {Ru}}, \bibinfo {author}
  {\bibfnamefont {T.-T.}\ \bibnamefont {Gao}}, \bibinfo {author} {\bibfnamefont
  {J.~M.}\ \bibnamefont {Moore}}, \ and\ \bibinfo {author} {\bibfnamefont
  {G.}~\bibnamefont {Yan}},\ }\bibfield  {title} {\enquote {\bibinfo {title}
  {Early predictor for the onset of critical transitions in networked dynamical
  systems},}\ }\href {\doibase 10.1103/PhysRevX.14.031009} {\bibfield
  {journal} {\bibinfo  {journal} {Phys. Rev. X}\ }\textbf {\bibinfo {volume}
  {14}},\ \bibinfo {pages} {031009} (\bibinfo {year} {2024})}\BibitemShut
  {NoStop}%
\bibitem [{\citenamefont {Harris}, \citenamefont {Gollo},\ and\ \citenamefont
  {Fulcher}(2024)}]{harris2024}%
  \BibitemOpen
  \bibfield  {author} {\bibinfo {author} {\bibfnamefont {B.}~\bibnamefont
  {Harris}}, \bibinfo {author} {\bibfnamefont {L.~L.}\ \bibnamefont {Gollo}}, \
  and\ \bibinfo {author} {\bibfnamefont {B.~D.}\ \bibnamefont {Fulcher}},\
  }\bibfield  {title} {\enquote {\bibinfo {title} {Tracking the distance to
  criticality in systems with unknown noise},}\ }\href {\doibase
  10.1103/PhysRevX.14.031021} {\bibfield  {journal} {\bibinfo  {journal} {Phys.
  Rev. X}\ }\textbf {\bibinfo {volume} {14}},\ \bibinfo {pages} {031021}
  (\bibinfo {year} {2024})}\BibitemShut {NoStop}%
\bibitem [{\citenamefont {Morr}\ and\ \citenamefont {Boers}(2024)}]{morr2024}%
  \BibitemOpen
  \bibfield  {author} {\bibinfo {author} {\bibfnamefont {A.}~\bibnamefont
  {Morr}}\ and\ \bibinfo {author} {\bibfnamefont {N.}~\bibnamefont {Boers}},\
  }\bibfield  {title} {\enquote {\bibinfo {title} {Detection of approaching
  critical transitions in natural systems driven by red noise},}\ }\href
  {\doibase 10.1103/PhysRevX.14.021037} {\bibfield  {journal} {\bibinfo
  {journal} {Phys. Rev. X}\ }\textbf {\bibinfo {volume} {14}},\ \bibinfo
  {pages} {021037} (\bibinfo {year} {2024})}\BibitemShut {NoStop}%
\bibitem [{\citenamefont {Evans}(1979)}]{evans1979}%
  \BibitemOpen
  \bibfield  {author} {\bibinfo {author} {\bibfnamefont {R.}~\bibnamefont
  {Evans}},\ }\bibfield  {title} {\enquote {\bibinfo {title} {The nature of the
  liquid-vapour interface and other topics in the statistical mechanics of
  non-uniform, classical fluids},}\ }\href {\doibase 10.1080/00018737900101365}
  {\bibfield  {journal} {\bibinfo  {journal} {Adv. Phys.}\ }\textbf {\bibinfo
  {volume} {28}},\ \bibinfo {pages} {143--200} (\bibinfo {year}
  {1979})}\BibitemShut {NoStop}%
\bibitem [{\citenamefont {Evans}(1992)}]{evans1992}%
  \BibitemOpen
  \bibfield  {author} {\bibinfo {author} {\bibfnamefont {R.}~\bibnamefont
  {Evans}},\ }\bibfield  {title} {\enquote {\bibinfo {title} {Density
  functionals in the theory of nonuniform fluids},}\ }in\ \href
  {https://books.google.com/books?id=-fNr2a4v3bYC&pg=PA85} {\emph {\bibinfo
  {booktitle} {Fundamentals of inhomogeneous fluids}}},\ \bibinfo {editor}
  {edited by\ \bibinfo {editor} {\bibfnamefont {D.}~\bibnamefont {Henderson}}}\
  (\bibinfo  {publisher} {Marcel Dekker},\ \bibinfo {address} {New York},\
  \bibinfo {year} {1992})\ Chap.~\bibinfo {chapter} {3}, pp.\ \bibinfo {pages}
  {85--176}\BibitemShut {NoStop}%
\bibitem [{\citenamefont {Evans}\ \emph {et~al.}(2016)\citenamefont {Evans},
  \citenamefont {Oettel}, \citenamefont {Roth},\ and\ \citenamefont
  {Kahl}}]{evans2016}%
  \BibitemOpen
  \bibfield  {author} {\bibinfo {author} {\bibfnamefont {R.}~\bibnamefont
  {Evans}}, \bibinfo {author} {\bibfnamefont {M.}~\bibnamefont {Oettel}},
  \bibinfo {author} {\bibfnamefont {R.}~\bibnamefont {Roth}}, \ and\ \bibinfo
  {author} {\bibfnamefont {G.}~\bibnamefont {Kahl}},\ }\bibfield  {title}
  {\enquote {\bibinfo {title} {New developments in classical density functional
  theory},}\ }\href {\doibase 10.1088/0953-8984/28/24/240401} {\bibfield
  {journal} {\bibinfo  {journal} {J. Phys.: Condens. Matter}\ }\textbf
  {\bibinfo {volume} {28}},\ \bibinfo {pages} {240401} (\bibinfo {year}
  {2016})}\BibitemShut {NoStop}%
\bibitem [{\citenamefont {{Santos-Silva}}\ \emph {et~al.}(2014)\citenamefont
  {{Santos-Silva}}, \citenamefont {Teixeira}, \citenamefont {{Anquetil-Deck}},\
  and\ \citenamefont {Cleaver}}]{teixera2014}%
  \BibitemOpen
  \bibfield  {author} {\bibinfo {author} {\bibfnamefont {T.}~\bibnamefont
  {{Santos-Silva}}}, \bibinfo {author} {\bibfnamefont {P.~I.~C.}\ \bibnamefont
  {Teixeira}}, \bibinfo {author} {\bibfnamefont {C.}~\bibnamefont
  {{Anquetil-Deck}}}, \ and\ \bibinfo {author} {\bibfnamefont {D.~J.}\
  \bibnamefont {Cleaver}},\ }\bibfield  {title} {\enquote {\bibinfo {title}
  {Neural-network approach to modeling liquid crystals in complex
  confinement},}\ }\href {\doibase 10.1103/PhysRevE.89.053316} {\bibfield
  {journal} {\bibinfo  {journal} {Phys. Rev. E}\ }\textbf {\bibinfo {volume}
  {89}},\ \bibinfo {pages} {053316} (\bibinfo {year} {2014})}\BibitemShut
  {NoStop}%
\bibitem [{\citenamefont {Lin}\ and\ \citenamefont {Oettel}(2019)}]{lin2019ml}%
  \BibitemOpen
  \bibfield  {author} {\bibinfo {author} {\bibfnamefont {S.-C.}\ \bibnamefont
  {Lin}}\ and\ \bibinfo {author} {\bibfnamefont {M.}~\bibnamefont {Oettel}},\
  }\bibfield  {title} {\enquote {\bibinfo {title} {A classical density
  functional from machine learning and a convolutional neural network},}\
  }\href {\doibase 10.21468/SciPostPhys.6.2.025} {\bibfield  {journal}
  {\bibinfo  {journal} {SciPost Phys.}\ }\textbf {\bibinfo {volume} {6}},\
  \bibinfo {pages} {025} (\bibinfo {year} {2019})}\BibitemShut {NoStop}%
\bibitem [{\citenamefont {Lin}, \citenamefont {Martius},\ and\ \citenamefont
  {Oettel}(2020)}]{lin2020ml}%
  \BibitemOpen
  \bibfield  {author} {\bibinfo {author} {\bibfnamefont {S.-C.}\ \bibnamefont
  {Lin}}, \bibinfo {author} {\bibfnamefont {G.}~\bibnamefont {Martius}}, \ and\
  \bibinfo {author} {\bibfnamefont {M.}~\bibnamefont {Oettel}},\ }\bibfield
  {title} {\enquote {\bibinfo {title} {Analytical classical density functionals
  from an equation learning network},}\ }\href {\doibase 10.1063/1.5135919}
  {\bibfield  {journal} {\bibinfo  {journal} {J. Chem. Phys.}\ }\textbf
  {\bibinfo {volume} {152}},\ \bibinfo {pages} {021102} (\bibinfo {year}
  {2020})}\BibitemShut {NoStop}%
\bibitem [{\citenamefont {Cats}\ \emph {et~al.}(2021)\citenamefont {Cats},
  \citenamefont {Kuipers}, \citenamefont {De~Wind}, \citenamefont {Van~Damme},
  \citenamefont {Coli}, \citenamefont {Dijkstra},\ and\ \citenamefont
  {Van~Roij}}]{cats2022ml}%
  \BibitemOpen
  \bibfield  {author} {\bibinfo {author} {\bibfnamefont {P.}~\bibnamefont
  {Cats}}, \bibinfo {author} {\bibfnamefont {S.}~\bibnamefont {Kuipers}},
  \bibinfo {author} {\bibfnamefont {S.}~\bibnamefont {De~Wind}}, \bibinfo
  {author} {\bibfnamefont {R.}~\bibnamefont {Van~Damme}}, \bibinfo {author}
  {\bibfnamefont {G.~M.}\ \bibnamefont {Coli}}, \bibinfo {author}
  {\bibfnamefont {M.}~\bibnamefont {Dijkstra}}, \ and\ \bibinfo {author}
  {\bibfnamefont {R.}~\bibnamefont {Van~Roij}},\ }\bibfield  {title} {\enquote
  {\bibinfo {title} {Machine-learning free-energy functionals using density
  profiles from simulations},}\ }\href {\doibase 10.1063/5.0042558} {\bibfield
  {journal} {\bibinfo  {journal} {APL Mater.}\ }\textbf {\bibinfo {volume}
  {9}},\ \bibinfo {pages} {031109} (\bibinfo {year} {2021})}\BibitemShut
  {NoStop}%
\bibitem [{\citenamefont {Yatsyshin}, \citenamefont {Kalliadasis},\ and\
  \citenamefont {Duncan}(2022)}]{yatsyshin2022}%
  \BibitemOpen
  \bibfield  {author} {\bibinfo {author} {\bibfnamefont {P.}~\bibnamefont
  {Yatsyshin}}, \bibinfo {author} {\bibfnamefont {S.}~\bibnamefont
  {Kalliadasis}}, \ and\ \bibinfo {author} {\bibfnamefont {A.~B.}\ \bibnamefont
  {Duncan}},\ }\bibfield  {title} {\enquote {\bibinfo {title}
  {Physics-constrained {{Bayesian}} inference of state functions in classical
  density-functional theory},}\ }\href {\doibase 10.1063/5.0071629} {\bibfield
  {journal} {\bibinfo  {journal} {J. Chem. Phys.}\ }\textbf {\bibinfo {volume}
  {156}},\ \bibinfo {pages} {074105} (\bibinfo {year} {2022})}\BibitemShut
  {NoStop}%
\bibitem [{\citenamefont {Fang}, \citenamefont {Gu},\ and\ \citenamefont
  {Wu}(2022)}]{fang2022}%
  \BibitemOpen
  \bibfield  {author} {\bibinfo {author} {\bibfnamefont {X.}~\bibnamefont
  {Fang}}, \bibinfo {author} {\bibfnamefont {M.}~\bibnamefont {Gu}}, \ and\
  \bibinfo {author} {\bibfnamefont {J.}~\bibnamefont {Wu}},\ }\bibfield
  {title} {\enquote {\bibinfo {title} {Reliable emulation of complex
  functionals by active learning with error control},}\ }\href {\doibase
  10.1063/5.0121805} {\bibfield  {journal} {\bibinfo  {journal} {J. Chem.
  Phys.}\ }\textbf {\bibinfo {volume} {157}},\ \bibinfo {pages} {214109}
  (\bibinfo {year} {2022})}\BibitemShut {NoStop}%
\bibitem [{\citenamefont {Qiao}\ \emph {et~al.}(2020)\citenamefont {Qiao},
  \citenamefont {Yu}, \citenamefont {Song}, \citenamefont {Zhao}, \citenamefont
  {Xu}, \citenamefont {Zhao},\ and\ \citenamefont {Gubbins}}]{qiao2020}%
  \BibitemOpen
  \bibfield  {author} {\bibinfo {author} {\bibfnamefont {C.}~\bibnamefont
  {Qiao}}, \bibinfo {author} {\bibfnamefont {X.}~\bibnamefont {Yu}}, \bibinfo
  {author} {\bibfnamefont {X.}~\bibnamefont {Song}}, \bibinfo {author}
  {\bibfnamefont {T.}~\bibnamefont {Zhao}}, \bibinfo {author} {\bibfnamefont
  {X.}~\bibnamefont {Xu}}, \bibinfo {author} {\bibfnamefont {S.}~\bibnamefont
  {Zhao}}, \ and\ \bibinfo {author} {\bibfnamefont {K.~E.}\ \bibnamefont
  {Gubbins}},\ }\bibfield  {title} {\enquote {\bibinfo {title} {Enhancing gas
  solubility in nanopores: A combined study using classical density functional
  theory and machine learning},}\ }\href {\doibase
  10.1021/acs.langmuir.0c01160} {\bibfield  {journal} {\bibinfo  {journal}
  {Langmuir}\ }\textbf {\bibinfo {volume} {36}},\ \bibinfo {pages} {8527--8536}
  (\bibinfo {year} {2020})}\BibitemShut {NoStop}%
\bibitem [{\citenamefont {Simon}\ \emph {et~al.}(2024)\citenamefont {Simon},
  \citenamefont {Weimar}, \citenamefont {Martius},\ and\ \citenamefont
  {Oettel}}]{simon2023mlPatchy}%
  \BibitemOpen
  \bibfield  {author} {\bibinfo {author} {\bibfnamefont {A.}~\bibnamefont
  {Simon}}, \bibinfo {author} {\bibfnamefont {J.}~\bibnamefont {Weimar}},
  \bibinfo {author} {\bibfnamefont {G.}~\bibnamefont {Martius}}, \ and\
  \bibinfo {author} {\bibfnamefont {M.}~\bibnamefont {Oettel}},\ }\bibfield
  {title} {\enquote {\bibinfo {title} {Machine learning of a density functional
  for anisotropic patchy particles},}\ }\href {\doibase
  10.1021/acs.jctc.3c01238} {\bibfield  {journal} {\bibinfo  {journal} {J.
  Chem. Theory Comput.}\ }\textbf {\bibinfo {volume} {20}},\ \bibinfo {pages}
  {1062--1077} (\bibinfo {year} {2024})}\BibitemShut {NoStop}%
\bibitem [{\citenamefont {Simon}\ \emph {et~al.}(2025)\citenamefont {Simon},
  \citenamefont {Belloni}, \citenamefont {Borgis},\ and\ \citenamefont
  {Oettel}}]{simon2024patchy}%
  \BibitemOpen
  \bibfield  {author} {\bibinfo {author} {\bibfnamefont {A.}~\bibnamefont
  {Simon}}, \bibinfo {author} {\bibfnamefont {L.}~\bibnamefont {Belloni}},
  \bibinfo {author} {\bibfnamefont {D.}~\bibnamefont {Borgis}}, \ and\ \bibinfo
  {author} {\bibfnamefont {M.}~\bibnamefont {Oettel}},\ }\bibfield  {title}
  {\enquote {\bibinfo {title} {The orientational structure of a model patchy
  particle fluid: {{Simulations}}, integral equations, density functional
  theory, and machine learning},}\ }\href {\doibase 10.1063/5.0248694}
  {\bibfield  {journal} {\bibinfo  {journal} {J. Chem. Phys.}\ }\textbf
  {\bibinfo {volume} {162}},\ \bibinfo {pages} {034503} (\bibinfo {year}
  {2025})}\BibitemShut {NoStop}%
\bibitem [{\citenamefont {Yang}\ \emph {et~al.}(2025)\citenamefont {Yang},
  \citenamefont {Pan}, \citenamefont {Sun},\ and\ \citenamefont
  {Wu}}]{yang2024}%
  \BibitemOpen
  \bibfield  {author} {\bibinfo {author} {\bibfnamefont {J.}~\bibnamefont
  {Yang}}, \bibinfo {author} {\bibfnamefont {R.}~\bibnamefont {Pan}}, \bibinfo
  {author} {\bibfnamefont {J.}~\bibnamefont {Sun}}, \ and\ \bibinfo {author}
  {\bibfnamefont {J.}~\bibnamefont {Wu}},\ }\bibfield  {title} {\enquote
  {\bibinfo {title} {High-dimensional operator learning for molecular density
  functional theory},}\ }\href {\doibase 10.1021/acs.jctc.5c00484} {\bibfield
  {journal} {\bibinfo  {journal} {J. Chem. Theory Comput.}\ ,\ \bibinfo {pages}
  {acs.jctc.5c00484}} (\bibinfo {year} {2025})}\BibitemShut {NoStop}%
\bibitem [{\citenamefont {Glitsch}, \citenamefont {Weimar},\ and\ \citenamefont
  {Oettel}(2025)}]{glitsch2025disks}%
  \BibitemOpen
  \bibfield  {author} {\bibinfo {author} {\bibfnamefont {F.}~\bibnamefont
  {Glitsch}}, \bibinfo {author} {\bibfnamefont {J.}~\bibnamefont {Weimar}}, \
  and\ \bibinfo {author} {\bibfnamefont {M.}~\bibnamefont {Oettel}},\
  }\bibfield  {title} {\enquote {\bibinfo {title} {Neural density functional
  theory in higher dimensions with convolutional layers},}\ }\href {\doibase
  10.1103/PhysRevE.111.055305} {\bibfield  {journal} {\bibinfo  {journal}
  {Phys. Rev. E}\ }\textbf {\bibinfo {volume} {111}},\ \bibinfo {pages}
  {055305} (\bibinfo {year} {2025})}\BibitemShut {NoStop}%
\bibitem [{\citenamefont {Dijkman}\ \emph {et~al.}(2025)\citenamefont
  {Dijkman}, \citenamefont {Dijkstra}, \citenamefont {Van~Roij}, \citenamefont
  {Welling}, \citenamefont {Van De~Meent},\ and\ \citenamefont
  {Ensing}}]{dijkman2024ml}%
  \BibitemOpen
  \bibfield  {author} {\bibinfo {author} {\bibfnamefont {J.}~\bibnamefont
  {Dijkman}}, \bibinfo {author} {\bibfnamefont {M.}~\bibnamefont {Dijkstra}},
  \bibinfo {author} {\bibfnamefont {R.}~\bibnamefont {Van~Roij}}, \bibinfo
  {author} {\bibfnamefont {M.}~\bibnamefont {Welling}}, \bibinfo {author}
  {\bibfnamefont {J.-W.}\ \bibnamefont {Van De~Meent}}, \ and\ \bibinfo
  {author} {\bibfnamefont {B.}~\bibnamefont {Ensing}},\ }\bibfield  {title}
  {\enquote {\bibinfo {title} {Learning neural free-energy functionals with
  pair-correlation matching},}\ }\href {\doibase
  10.1103/PhysRevLett.134.056103} {\bibfield  {journal} {\bibinfo  {journal}
  {Phys. Rev. Lett.}\ }\textbf {\bibinfo {volume} {134}},\ \bibinfo {pages}
  {056103} (\bibinfo {year} {2025})}\BibitemShut {NoStop}%
\bibitem [{\citenamefont {Kelley}\ \emph {et~al.}(2024)\citenamefont {Kelley},
  \citenamefont {Quinton}, \citenamefont {Fazel}, \citenamefont {Karimitari},
  \citenamefont {Sutton},\ and\ \citenamefont {Sundararaman}}]{kelley2024ml}%
  \BibitemOpen
  \bibfield  {author} {\bibinfo {author} {\bibfnamefont {M.~M.}\ \bibnamefont
  {Kelley}}, \bibinfo {author} {\bibfnamefont {J.}~\bibnamefont {Quinton}},
  \bibinfo {author} {\bibfnamefont {K.}~\bibnamefont {Fazel}}, \bibinfo
  {author} {\bibfnamefont {N.}~\bibnamefont {Karimitari}}, \bibinfo {author}
  {\bibfnamefont {C.}~\bibnamefont {Sutton}}, \ and\ \bibinfo {author}
  {\bibfnamefont {R.}~\bibnamefont {Sundararaman}},\ }\bibfield  {title}
  {\enquote {\bibinfo {title} {Bridging electronic and classical
  density-functional theory using universal machine-learned functional
  approximations},}\ }\href {\doibase 10.1063/5.0223792} {\bibfield  {journal}
  {\bibinfo  {journal} {J. Chem. Phys.}\ }\textbf {\bibinfo {volume} {161}},\
  \bibinfo {pages} {144101} (\bibinfo {year} {2024})}\BibitemShut {NoStop}%
\bibitem [{\citenamefont {{Malpica-Morales}}\ \emph {et~al.}(2023)\citenamefont
  {{Malpica-Morales}}, \citenamefont {Yatsyshin}, \citenamefont
  {{Dur{\'a}n-Olivencia}},\ and\ \citenamefont
  {Kalliadasis}}]{malpica-morales2023}%
  \BibitemOpen
  \bibfield  {author} {\bibinfo {author} {\bibfnamefont {A.}~\bibnamefont
  {{Malpica-Morales}}}, \bibinfo {author} {\bibfnamefont {P.}~\bibnamefont
  {Yatsyshin}}, \bibinfo {author} {\bibfnamefont {M.~A.}\ \bibnamefont
  {{Dur{\'a}n-Olivencia}}}, \ and\ \bibinfo {author} {\bibfnamefont
  {S.}~\bibnamefont {Kalliadasis}},\ }\bibfield  {title} {\enquote {\bibinfo
  {title} {Physics-informed {{Bayesian}} inference of external potentials in
  classical density-functional theory},}\ }\href {\doibase 10.1063/5.0146920}
  {\bibfield  {journal} {\bibinfo  {journal} {J. Chem. Phys.}\ }\textbf
  {\bibinfo {volume} {159}},\ \bibinfo {pages} {104109} (\bibinfo {year}
  {2023})}\BibitemShut {NoStop}%
\bibitem [{\citenamefont {Pan}\ \emph {et~al.}(2025)\citenamefont {Pan},
  \citenamefont {Fang}, \citenamefont {Azizzadenesheli}, \citenamefont
  {{Liu-Schiaffini}}, \citenamefont {Gu},\ and\ \citenamefont {Wu}}]{pan2025}%
  \BibitemOpen
  \bibfield  {author} {\bibinfo {author} {\bibfnamefont {R.}~\bibnamefont
  {Pan}}, \bibinfo {author} {\bibfnamefont {X.}~\bibnamefont {Fang}}, \bibinfo
  {author} {\bibfnamefont {K.}~\bibnamefont {Azizzadenesheli}}, \bibinfo
  {author} {\bibfnamefont {M.}~\bibnamefont {{Liu-Schiaffini}}}, \bibinfo
  {author} {\bibfnamefont {M.}~\bibnamefont {Gu}}, \ and\ \bibinfo {author}
  {\bibfnamefont {J.}~\bibnamefont {Wu}},\ }\href {\doibase
  10.48550/ARXIV.2506.06623} {\enquote {\bibinfo {title} {Neural operators for
  forward and inverse potential-density mappings in classical density
  functional theory},}\ } (\bibinfo {year} {2025}),\ \Eprint
  {http://arxiv.org/abs/2506.06623} {arXiv:2506.06623 [physics.chem-ph]}
  \BibitemShut {NoStop}%
\bibitem [{\citenamefont {Bui}\ and\ \citenamefont {Cox}(2024)}]{bui2024}%
  \BibitemOpen
  \bibfield  {author} {\bibinfo {author} {\bibfnamefont {A.~T.}\ \bibnamefont
  {Bui}}\ and\ \bibinfo {author} {\bibfnamefont {S.~J.}\ \bibnamefont {Cox}},\
  }\bibfield  {title} {\enquote {\bibinfo {title} {A classical density
  functional theory for solvation across length scales},}\ }\href {\doibase
  10.1063/5.0223750} {\bibfield  {journal} {\bibinfo  {journal} {J. Chem.
  Phys.}\ }\textbf {\bibinfo {volume} {161}},\ \bibinfo {pages} {104103}
  (\bibinfo {year} {2024})}\BibitemShut {NoStop}%
\bibitem [{\citenamefont {Bui}\ and\ \citenamefont
  {Cox}(2025{\natexlab{a}})}]{bui2025electromechanics}%
  \BibitemOpen
  \bibfield  {author} {\bibinfo {author} {\bibfnamefont {A.~T.}\ \bibnamefont
  {Bui}}\ and\ \bibinfo {author} {\bibfnamefont {S.~J.}\ \bibnamefont {Cox}},\
  }\bibfield  {title} {\enquote {\bibinfo {title} {A first principles approach
  to electromechanics in liquids},}\ }\href {\doibase 10.1088/1361-648x/ade7e7}
  {\bibfield  {journal} {\bibinfo  {journal} {J. Phys.: Condens. Matter}\ }
  (\bibinfo {year} {2025}{\natexlab{a}}),\
  10.1088/1361-648x/ade7e7}\BibitemShut {NoStop}%
\bibitem [{\citenamefont {Bui}\ and\ \citenamefont
  {Cox}(2025{\natexlab{b}})}]{bui2025dielectrocapillarity}%
  \BibitemOpen
  \bibfield  {author} {\bibinfo {author} {\bibfnamefont {A.~T.}\ \bibnamefont
  {Bui}}\ and\ \bibinfo {author} {\bibfnamefont {S.~J.}\ \bibnamefont {Cox}},\
  }\href {\doibase 10.48550/ARXIV.2503.09855} {\enquote {\bibinfo {title}
  {Dielectrocapillarity for exquisite control of fluids},}\ } (\bibinfo {year}
  {2025}{\natexlab{b}}),\ \Eprint {http://arxiv.org/abs/2503.09855}
  {arXiv:2503.09855 [cond-mat.soft]} \BibitemShut {NoStop}%
\bibitem [{\citenamefont {Bui}\ and\ \citenamefont
  {Cox}(2025{\natexlab{c}})}]{bui2024neuralrpm}%
  \BibitemOpen
  \bibfield  {author} {\bibinfo {author} {\bibfnamefont {A.~T.}\ \bibnamefont
  {Bui}}\ and\ \bibinfo {author} {\bibfnamefont {S.~J.}\ \bibnamefont {Cox}},\
  }\bibfield  {title} {\enquote {\bibinfo {title} {Learning classical density
  functionals for ionic fluids},}\ }\href {\doibase
  10.1103/PhysRevLett.134.148001} {\bibfield  {journal} {\bibinfo  {journal}
  {Phys. Rev. Lett.}\ }\textbf {\bibinfo {volume} {134}},\ \bibinfo {pages}
  {148001} (\bibinfo {year} {2025}{\natexlab{c}})}\BibitemShut {NoStop}%
\bibitem [{\citenamefont {Samm{\"u}ller}\ \emph {et~al.}(2023)\citenamefont
  {Samm{\"u}ller}, \citenamefont {Hermann}, \citenamefont {de~las Heras},\ and\
  \citenamefont {Schmidt}}]{sammueller2023neural}%
  \BibitemOpen
  \bibfield  {author} {\bibinfo {author} {\bibfnamefont {F.}~\bibnamefont
  {Samm{\"u}ller}}, \bibinfo {author} {\bibfnamefont {S.}~\bibnamefont
  {Hermann}}, \bibinfo {author} {\bibfnamefont {D.}~\bibnamefont {de~las
  Heras}}, \ and\ \bibinfo {author} {\bibfnamefont {M.}~\bibnamefont
  {Schmidt}},\ }\bibfield  {title} {\enquote {\bibinfo {title} {Neural
  functional theory for inhomogeneous fluids: {{Fundamentals}} and
  applications},}\ }\href {\doibase 10.1073/pnas.2312484120} {\bibfield
  {journal} {\bibinfo  {journal} {Proc. Natl. Acad. Sci. U.S.A.}\ }\textbf
  {\bibinfo {volume} {120}},\ \bibinfo {pages} {e2312484120} (\bibinfo {year}
  {2023})}\BibitemShut {NoStop}%
\bibitem [{\citenamefont {Samm{\"u}ller}, \citenamefont {Hermann},\ and\
  \citenamefont {Schmidt}(2024)}]{sammueller2023whyNeural}%
  \BibitemOpen
  \bibfield  {author} {\bibinfo {author} {\bibfnamefont {F.}~\bibnamefont
  {Samm{\"u}ller}}, \bibinfo {author} {\bibfnamefont {S.}~\bibnamefont
  {Hermann}}, \ and\ \bibinfo {author} {\bibfnamefont {M.}~\bibnamefont
  {Schmidt}},\ }\bibfield  {title} {\enquote {\bibinfo {title} {Why neural
  functionals suit statistical mechanics},}\ }\href {\doibase
  10.1088/1361-648X/ad326f} {\bibfield  {journal} {\bibinfo  {journal} {J.
  Phys.: Condens. Matter}\ }\textbf {\bibinfo {volume} {36}},\ \bibinfo {pages}
  {243002} (\bibinfo {year} {2024})}\BibitemShut {NoStop}%
\bibitem [{\citenamefont {Samm{\"u}ller}()}]{sammueller2023neuralTutorial}%
  \BibitemOpen
  \bibfield  {author} {\bibinfo {author} {\bibfnamefont {F.}~\bibnamefont
  {Samm{\"u}ller}},\ }\href@noop {} {\enquote {\bibinfo {title} {Neural
  functional theory for inhomogeneous fluids -- {{Tutorial}}},}\ }\bibinfo
  {note} {\url{https://github.com/sfalmo/NeuralDFT-Tutorial}}\BibitemShut
  {NoStop}%
\bibitem [{\citenamefont {de~las Heras}\ \emph {et~al.}(2023)\citenamefont
  {de~las Heras}, \citenamefont {Zimmermann}, \citenamefont {Samm{\"u}ller},
  \citenamefont {Hermann},\ and\ \citenamefont
  {Schmidt}}]{delasheras2023perspective}%
  \BibitemOpen
  \bibfield  {author} {\bibinfo {author} {\bibfnamefont {D.}~\bibnamefont
  {de~las Heras}}, \bibinfo {author} {\bibfnamefont {T.}~\bibnamefont
  {Zimmermann}}, \bibinfo {author} {\bibfnamefont {F.}~\bibnamefont
  {Samm{\"u}ller}}, \bibinfo {author} {\bibfnamefont {S.}~\bibnamefont
  {Hermann}}, \ and\ \bibinfo {author} {\bibfnamefont {M.}~\bibnamefont
  {Schmidt}},\ }\bibfield  {title} {\enquote {\bibinfo {title} {Perspective:
  {{How}} to overcome dynamical density functional theory},}\ }\href {\doibase
  10.1088/1361-648X/accb33} {\bibfield  {journal} {\bibinfo  {journal} {J.
  Phys.: Condens. Matter}\ }\textbf {\bibinfo {volume} {35}},\ \bibinfo {pages}
  {271501} (\bibinfo {year} {2023})}\BibitemShut {NoStop}%
\bibitem [{\citenamefont {Zimmermann}\ \emph {et~al.}(2024)\citenamefont
  {Zimmermann}, \citenamefont {Samm{\"u}ller}, \citenamefont {Hermann},
  \citenamefont {Schmidt},\ and\ \citenamefont {de~las
  Heras}}]{zimmermann2024ml}%
  \BibitemOpen
  \bibfield  {author} {\bibinfo {author} {\bibfnamefont {T.}~\bibnamefont
  {Zimmermann}}, \bibinfo {author} {\bibfnamefont {F.}~\bibnamefont
  {Samm{\"u}ller}}, \bibinfo {author} {\bibfnamefont {S.}~\bibnamefont
  {Hermann}}, \bibinfo {author} {\bibfnamefont {M.}~\bibnamefont {Schmidt}}, \
  and\ \bibinfo {author} {\bibfnamefont {D.}~\bibnamefont {de~las Heras}},\
  }\bibfield  {title} {\enquote {\bibinfo {title} {Neural force functional for
  non-equilibrium many-body colloidal systems},}\ }\href {\doibase
  10.1088/2632-2153/ad7191} {\bibfield  {journal} {\bibinfo  {journal} {Mach.
  Learn.: Sci. Technol.}\ }\textbf {\bibinfo {volume} {5}},\ \bibinfo {pages}
  {035062} (\bibinfo {year} {2024})}\BibitemShut {NoStop}%
\bibitem [{\citenamefont {Samm{\"u}ller}\ \emph {et~al.}(2024)\citenamefont
  {Samm{\"u}ller}, \citenamefont {Robitschko}, \citenamefont {Hermann},\ and\
  \citenamefont {Schmidt}}]{sammueller2024hyperDFT}%
  \BibitemOpen
  \bibfield  {author} {\bibinfo {author} {\bibfnamefont {F.}~\bibnamefont
  {Samm{\"u}ller}}, \bibinfo {author} {\bibfnamefont {S.}~\bibnamefont
  {Robitschko}}, \bibinfo {author} {\bibfnamefont {S.}~\bibnamefont {Hermann}},
  \ and\ \bibinfo {author} {\bibfnamefont {M.}~\bibnamefont {Schmidt}},\
  }\bibfield  {title} {\enquote {\bibinfo {title} {Hyperdensity functional
  theory of soft matter},}\ }\href {\doibase 10.1103/PhysRevLett.133.098201}
  {\bibfield  {journal} {\bibinfo  {journal} {Phys. Rev. Lett.}\ }\textbf
  {\bibinfo {volume} {133}},\ \bibinfo {pages} {098201} (\bibinfo {year}
  {2024})}\BibitemShut {NoStop}%
\bibitem [{\citenamefont {Samm{\"u}ller}\ and\ \citenamefont
  {Schmidt}(2025)}]{sammueller2024whyhyperDFT}%
  \BibitemOpen
  \bibfield  {author} {\bibinfo {author} {\bibfnamefont {F.}~\bibnamefont
  {Samm{\"u}ller}}\ and\ \bibinfo {author} {\bibfnamefont {M.}~\bibnamefont
  {Schmidt}},\ }\bibfield  {title} {\enquote {\bibinfo {title} {Why
  hyperdensity functionals describe any equilibrium observable},}\ }\href
  {\doibase 10.1088/1361-648X/ad98da} {\bibfield  {journal} {\bibinfo
  {journal} {J. Phys.: Condens. Matter}\ }\textbf {\bibinfo {volume} {37}},\
  \bibinfo {pages} {083001} (\bibinfo {year} {2025})}\BibitemShut {NoStop}%
\bibitem [{\citenamefont {Samm{\"u}ller}\ and\ \citenamefont
  {Schmidt}(2024)}]{sammueller2024pairmatching}%
  \BibitemOpen
  \bibfield  {author} {\bibinfo {author} {\bibfnamefont {F.}~\bibnamefont
  {Samm{\"u}ller}}\ and\ \bibinfo {author} {\bibfnamefont {M.}~\bibnamefont
  {Schmidt}},\ }\bibfield  {title} {\enquote {\bibinfo {title} {Neural density
  functionals: {{Local}} learning and pair-correlation matching},}\ }\href
  {\doibase 10.1103/PhysRevE.110.L032601} {\bibfield  {journal} {\bibinfo
  {journal} {Phys. Rev. E}\ }\textbf {\bibinfo {volume} {110}},\ \bibinfo
  {pages} {L032601} (\bibinfo {year} {2024})}\BibitemShut {NoStop}%
\bibitem [{\citenamefont {Samm{\"u}ller}, \citenamefont {Schmidt},\ and\
  \citenamefont {Evans}(2025)}]{sammueller2024attraction}%
  \BibitemOpen
  \bibfield  {author} {\bibinfo {author} {\bibfnamefont {F.}~\bibnamefont
  {Samm{\"u}ller}}, \bibinfo {author} {\bibfnamefont {M.}~\bibnamefont
  {Schmidt}}, \ and\ \bibinfo {author} {\bibfnamefont {R.}~\bibnamefont
  {Evans}},\ }\bibfield  {title} {\enquote {\bibinfo {title} {Neural density
  functional theory of liquid-gas phase coexistence},}\ }\href {\doibase
  10.1103/PhysRevX.15.011013} {\bibfield  {journal} {\bibinfo  {journal} {Phys.
  Rev. X}\ }\textbf {\bibinfo {volume} {15}},\ \bibinfo {pages} {011013}
  (\bibinfo {year} {2025})}\BibitemShut {NoStop}%
\bibitem [{\citenamefont {Buchanan}(2025)}]{buchanan2025attraction}%
  \BibitemOpen
  \bibfield  {author} {\bibinfo {author} {\bibfnamefont {M.}~\bibnamefont
  {Buchanan}},\ }\bibfield  {title} {\enquote {\bibinfo {title} {Machine
  learning predicts liquid--gas transition},}\ }\href {\doibase
  10.1103/Physics.18.17} {\bibfield  {journal} {\bibinfo  {journal} {Physics}\
  }\textbf {\bibinfo {volume} {18}},\ \bibinfo {pages} {17} (\bibinfo {year}
  {2025})}\BibitemShut {NoStop}%
\bibitem [{\citenamefont {Kampa}\ \emph {et~al.}(2025)\citenamefont {Kampa},
  \citenamefont {Samm{\"u}ller}, \citenamefont {Schmidt},\ and\ \citenamefont
  {Evans}}]{kampa2024meta}%
  \BibitemOpen
  \bibfield  {author} {\bibinfo {author} {\bibfnamefont {S.~M.}\ \bibnamefont
  {Kampa}}, \bibinfo {author} {\bibfnamefont {F.}~\bibnamefont
  {Samm{\"u}ller}}, \bibinfo {author} {\bibfnamefont {M.}~\bibnamefont
  {Schmidt}}, \ and\ \bibinfo {author} {\bibfnamefont {R.}~\bibnamefont
  {Evans}},\ }\bibfield  {title} {\enquote {\bibinfo {title} {Metadensity
  functional theory for classical fluids: {{Extracting}} the pair potential},}\
  }\href {\doibase 10.1103/PhysRevLett.134.107301} {\bibfield  {journal}
  {\bibinfo  {journal} {Phys. Rev. Lett.}\ }\textbf {\bibinfo {volume} {134}},\
  \bibinfo {pages} {107301} (\bibinfo {year} {2025})}\BibitemShut {NoStop}%
\bibitem [{\citenamefont {Telo Da~Gama}\ and\ \citenamefont
  {Evans}(1983{\natexlab{a}})}]{telodagama1983one}%
  \BibitemOpen
  \bibfield  {author} {\bibinfo {author} {\bibfnamefont {M.}~\bibnamefont {Telo
  Da~Gama}}\ and\ \bibinfo {author} {\bibfnamefont {R.}~\bibnamefont {Evans}},\
  }\bibfield  {title} {\enquote {\bibinfo {title} {The structure and surface
  tension of the liquid-vapour interface near the upper critical end point of a
  binary mixture of {{Lennard-Jones}} fluids: {{I}}. {{The}} two phase
  region},}\ }\href {\doibase 10.1080/00268978300100181} {\bibfield  {journal}
  {\bibinfo  {journal} {Mol. Phys.}\ }\textbf {\bibinfo {volume} {48}},\
  \bibinfo {pages} {229--250} (\bibinfo {year}
  {1983}{\natexlab{a}})}\BibitemShut {NoStop}%
\bibitem [{\citenamefont {Telo Da~Gama}\ and\ \citenamefont
  {Evans}(1983{\natexlab{b}})}]{telodagama1983two}%
  \BibitemOpen
  \bibfield  {author} {\bibinfo {author} {\bibfnamefont {M.}~\bibnamefont {Telo
  Da~Gama}}\ and\ \bibinfo {author} {\bibfnamefont {R.}~\bibnamefont {Evans}},\
  }\bibfield  {title} {\enquote {\bibinfo {title} {The structure and surface
  tension of the liquid-vapour interface near the upper critical end point of a
  binary mixture of {{Lennard-Jones}} fluids: {{II}}. {{The}} three phase
  region and the {{Cahn}} wetting transition},}\ }\href {\doibase
  10.1080/00268978300100191} {\bibfield  {journal} {\bibinfo  {journal} {Mol.
  Phys.}\ }\textbf {\bibinfo {volume} {48}},\ \bibinfo {pages} {251--266}
  (\bibinfo {year} {1983}{\natexlab{b}})}\BibitemShut {NoStop}%
\bibitem [{\citenamefont {Tarazona}, \citenamefont {Evans},\ and\ \citenamefont
  {Marini Bettolo~Marconi}(1985)}]{tarazona1985interface}%
  \BibitemOpen
  \bibfield  {author} {\bibinfo {author} {\bibfnamefont {P.}~\bibnamefont
  {Tarazona}}, \bibinfo {author} {\bibfnamefont {R.}~\bibnamefont {Evans}}, \
  and\ \bibinfo {author} {\bibfnamefont {U.}~\bibnamefont {Marini
  Bettolo~Marconi}},\ }\bibfield  {title} {\enquote {\bibinfo {title} {Pairwise
  correlations at a fluid-fluid interface: {{The}} influence of a wetting
  film},}\ }\href {\doibase 10.1080/00268978500101051} {\bibfield  {journal}
  {\bibinfo  {journal} {Mol. Phys.}\ }\textbf {\bibinfo {volume} {54}},\
  \bibinfo {pages} {1357--1392} (\bibinfo {year} {1985})}\BibitemShut {NoStop}%
\bibitem [{\citenamefont {Hadjiagapiou}\ and\ \citenamefont
  {Evans}(1985)}]{hadjiagapiou1985}%
  \BibitemOpen
  \bibfield  {author} {\bibinfo {author} {\bibfnamefont {I.}~\bibnamefont
  {Hadjiagapiou}}\ and\ \bibinfo {author} {\bibfnamefont {R.}~\bibnamefont
  {Evans}},\ }\bibfield  {title} {\enquote {\bibinfo {title} {Adsorption from a
  binary fluid mixture: {{The}} composite wetting film at the solid-vapour
  interface},}\ }\href {\doibase 10.1080/00268978500100301} {\bibfield
  {journal} {\bibinfo  {journal} {Mol. Phys.}\ }\textbf {\bibinfo {volume}
  {54}},\ \bibinfo {pages} {383--406} (\bibinfo {year} {1985})}\BibitemShut
  {NoStop}%
\bibitem [{\citenamefont {Schmid}\ and\ \citenamefont
  {Wilding}(2001)}]{schmid2001wetting}%
  \BibitemOpen
  \bibfield  {author} {\bibinfo {author} {\bibfnamefont {F.}~\bibnamefont
  {Schmid}}\ and\ \bibinfo {author} {\bibfnamefont {N.~B.}\ \bibnamefont
  {Wilding}},\ }\bibfield  {title} {\enquote {\bibinfo {title} {Wetting of a
  symmetrical binary fluid mixture on a wall},}\ }\href {\doibase
  10.1103/PhysRevE.63.031201} {\bibfield  {journal} {\bibinfo  {journal} {Phys.
  Rev. E}\ }\textbf {\bibinfo {volume} {63}},\ \bibinfo {pages} {031201}
  (\bibinfo {year} {2001})}\BibitemShut {NoStop}%
\bibitem [{\citenamefont {Wilding}\ and\ \citenamefont
  {Schmid}(2002)}]{wilding2002}%
  \BibitemOpen
  \bibfield  {author} {\bibinfo {author} {\bibfnamefont {N.}~\bibnamefont
  {Wilding}}\ and\ \bibinfo {author} {\bibfnamefont {F.}~\bibnamefont
  {Schmid}},\ }\bibfield  {title} {\enquote {\bibinfo {title} {Wetting of a
  symmetrical binary fluid mixture on a wall},}\ }\href {\doibase
  10.1016/S0010-4655(02)00234-5} {\bibfield  {journal} {\bibinfo  {journal}
  {Comput. Phys. Commun.}\ }\textbf {\bibinfo {volume} {147}},\ \bibinfo
  {pages} {149--153} (\bibinfo {year} {2002})}\BibitemShut {NoStop}%
\bibitem [{\citenamefont {Napari}\ \emph {et~al.}(1999)\citenamefont {Napari},
  \citenamefont {Laaksonen}, \citenamefont {Talanquer},\ and\ \citenamefont
  {Oxtoby}}]{napari1999}%
  \BibitemOpen
  \bibfield  {author} {\bibinfo {author} {\bibfnamefont {I.}~\bibnamefont
  {Napari}}, \bibinfo {author} {\bibfnamefont {A.}~\bibnamefont {Laaksonen}},
  \bibinfo {author} {\bibfnamefont {V.}~\bibnamefont {Talanquer}}, \ and\
  \bibinfo {author} {\bibfnamefont {D.~W.}\ \bibnamefont {Oxtoby}},\ }\bibfield
   {title} {\enquote {\bibinfo {title} {A density functional study of
  liquid--liquid interfaces in partially miscible systems},}\ }\href {\doibase
  10.1063/1.478490} {\bibfield  {journal} {\bibinfo  {journal} {J. Chem.
  Phys.}\ }\textbf {\bibinfo {volume} {110}},\ \bibinfo {pages} {5906--5912}
  (\bibinfo {year} {1999})}\BibitemShut {NoStop}%
\bibitem [{\citenamefont {Rowlinson}\ and\ \citenamefont
  {Widom}(2002)}]{rowlinsonwidombook}%
  \BibitemOpen
  \bibfield  {author} {\bibinfo {author} {\bibfnamefont {J.~S.}\ \bibnamefont
  {Rowlinson}}\ and\ \bibinfo {author} {\bibfnamefont {B.}~\bibnamefont
  {Widom}},\ }\href@noop {} {\emph {\bibinfo {title} {Molecular Theory of
  Capillarity}}}\ (\bibinfo  {publisher} {Dover Publications},\ \bibinfo
  {address} {Mineola, New York},\ \bibinfo {year} {2002})\BibitemShut {NoStop}%
\bibitem [{\citenamefont {{Sch{\"o}ll-Paschinger}}\ and\ \citenamefont
  {Kahl}(2003)}]{schoellpaschinger2003}%
  \BibitemOpen
  \bibfield  {author} {\bibinfo {author} {\bibfnamefont {E.}~\bibnamefont
  {{Sch{\"o}ll-Paschinger}}}\ and\ \bibinfo {author} {\bibfnamefont
  {G.}~\bibnamefont {Kahl}},\ }\bibfield  {title} {\enquote {\bibinfo {title}
  {Self-consistent {{Ornstein}}--{{Zernike}} approximation for a binary
  symmetric fluid mixture},}\ }\href {\doibase 10.1063/1.1557053} {\bibfield
  {journal} {\bibinfo  {journal} {J. Chem. Phys.}\ }\textbf {\bibinfo {volume}
  {118}},\ \bibinfo {pages} {7414--7424} (\bibinfo {year} {2003})}\BibitemShut
  {NoStop}%
\bibitem [{\citenamefont {Antonevych}, \citenamefont {Forstmann},\ and\
  \citenamefont {{Diaz-Herrera}}(2002)}]{antonevych2002}%
  \BibitemOpen
  \bibfield  {author} {\bibinfo {author} {\bibfnamefont {O.}~\bibnamefont
  {Antonevych}}, \bibinfo {author} {\bibfnamefont {F.}~\bibnamefont
  {Forstmann}}, \ and\ \bibinfo {author} {\bibfnamefont {E.}~\bibnamefont
  {{Diaz-Herrera}}},\ }\bibfield  {title} {\enquote {\bibinfo {title} {Phase
  diagram of symmetric binary fluid mixtures: {{First-order}} or second-order
  demixing},}\ }\href {\doibase 10.1103/PhysRevE.65.061504} {\bibfield
  {journal} {\bibinfo  {journal} {Phys. Rev. E}\ }\textbf {\bibinfo {volume}
  {65}},\ \bibinfo {pages} {061504} (\bibinfo {year} {2002})}\BibitemShut
  {NoStop}%
\bibitem [{\citenamefont {{Mart{\'i}nez-Ruiz}}, \citenamefont {{Moreno-Ventas
  Bravo}},\ and\ \citenamefont {Blas}(2015)}]{martinezruiz2015}%
  \BibitemOpen
  \bibfield  {author} {\bibinfo {author} {\bibfnamefont {F.~J.}\ \bibnamefont
  {{Mart{\'i}nez-Ruiz}}}, \bibinfo {author} {\bibfnamefont {A.~I.}\
  \bibnamefont {{Moreno-Ventas Bravo}}}, \ and\ \bibinfo {author}
  {\bibfnamefont {F.~J.}\ \bibnamefont {Blas}},\ }\bibfield  {title} {\enquote
  {\bibinfo {title} {Liquid-liquid interfacial properties of a symmetrical
  {{Lennard-Jones}} binary mixture},}\ }\href {\doibase 10.1063/1.4930276}
  {\bibfield  {journal} {\bibinfo  {journal} {J. Chem. Phys.}\ }\textbf
  {\bibinfo {volume} {143}},\ \bibinfo {pages} {104706} (\bibinfo {year}
  {2015})}\BibitemShut {NoStop}%
\bibitem [{\citenamefont {Roy}\ and\ \citenamefont
  {H\"{o}fling}(2024)}]{roy2024}%
  \BibitemOpen
  \bibfield  {author} {\bibinfo {author} {\bibfnamefont {S.}~\bibnamefont
  {Roy}}\ and\ \bibinfo {author} {\bibfnamefont {F.}~\bibnamefont
  {H\"{o}fling}},\ }\bibfield  {title} {\enquote {\bibinfo {title} {Critical
  surface adsorption of confined binary liquids with locally conserved mass and
  composition},}\ }\href {\doibase 10.1080/00268976.2024.2391998} {\bibfield
  {journal} {\bibinfo  {journal} {Mol. Phys.}\ }\textbf {\bibinfo {volume}
  {122}},\ \bibinfo {pages} {e2391998} (\bibinfo {year} {2024})}\BibitemShut
  {NoStop}%
\bibitem [{\citenamefont {Wilding}(1997)}]{wilding1997}%
  \BibitemOpen
  \bibfield  {author} {\bibinfo {author} {\bibfnamefont {N.~B.}\ \bibnamefont
  {Wilding}},\ }\bibfield  {title} {\enquote {\bibinfo {title} {Critical end
  point behavior in a binary fluid mixture},}\ }\href {\doibase
  10.1103/PhysRevE.55.6624} {\bibfield  {journal} {\bibinfo  {journal} {Phys.
  Rev. E}\ }\textbf {\bibinfo {volume} {55}},\ \bibinfo {pages} {6624--6631}
  (\bibinfo {year} {1997})}\BibitemShut {NoStop}%
\bibitem [{\citenamefont {Wilding}, \citenamefont {Schmid},\ and\ \citenamefont
  {Nielaba}(1998)}]{wilding1998}%
  \BibitemOpen
  \bibfield  {author} {\bibinfo {author} {\bibfnamefont {N.~B.}\ \bibnamefont
  {Wilding}}, \bibinfo {author} {\bibfnamefont {F.}~\bibnamefont {Schmid}}, \
  and\ \bibinfo {author} {\bibfnamefont {P.}~\bibnamefont {Nielaba}},\
  }\bibfield  {title} {\enquote {\bibinfo {title} {Liquid-vapor phase behavior
  of a symmetrical binary fluid mixture},}\ }\href {\doibase
  10.1103/PhysRevE.58.2201} {\bibfield  {journal} {\bibinfo  {journal} {Phys.
  Rev. E}\ }\textbf {\bibinfo {volume} {58}},\ \bibinfo {pages} {2201--2212}
  (\bibinfo {year} {1998})}\BibitemShut {NoStop}%
\bibitem [{\citenamefont {K{\"o}finger}, \citenamefont {Kahl},\ and\
  \citenamefont {Wilding}(2006)}]{koefinger2006epl}%
  \BibitemOpen
  \bibfield  {author} {\bibinfo {author} {\bibfnamefont {J.}~\bibnamefont
  {K{\"o}finger}}, \bibinfo {author} {\bibfnamefont {G.}~\bibnamefont {Kahl}},
  \ and\ \bibinfo {author} {\bibfnamefont {N.~B.}\ \bibnamefont {Wilding}},\
  }\bibfield  {title} {\enquote {\bibinfo {title} {Phase behaviour of a
  symmetrical binary mixture in a field},}\ }\href {\doibase
  10.1209/epl/i2006-10087-7} {\bibfield  {journal} {\bibinfo  {journal}
  {Europhys. Lett.}\ }\textbf {\bibinfo {volume} {75}},\ \bibinfo {pages}
  {234--240} (\bibinfo {year} {2006})}\BibitemShut {NoStop}%
\bibitem [{\citenamefont {K{\"o}finger}, \citenamefont {Wilding},\ and\
  \citenamefont {Kahl}(2006)}]{koefinger2006jcp}%
  \BibitemOpen
  \bibfield  {author} {\bibinfo {author} {\bibfnamefont {J.}~\bibnamefont
  {K{\"o}finger}}, \bibinfo {author} {\bibfnamefont {N.~B.}\ \bibnamefont
  {Wilding}}, \ and\ \bibinfo {author} {\bibfnamefont {G.}~\bibnamefont
  {Kahl}},\ }\bibfield  {title} {\enquote {\bibinfo {title} {Phase behavior of
  a symmetrical binary fluid mixture},}\ }\href {\doibase 10.1063/1.2393241}
  {\bibfield  {journal} {\bibinfo  {journal} {J. Chem. Phys.}\ }\textbf
  {\bibinfo {volume} {125}},\ \bibinfo {pages} {234503} (\bibinfo {year}
  {2006})}\BibitemShut {NoStop}%
\bibitem [{\citenamefont {Roy}, \citenamefont {Dietrich},\ and\ \citenamefont
  {H\"{o}fling}(2016)}]{roy2016}%
  \BibitemOpen
  \bibfield  {author} {\bibinfo {author} {\bibfnamefont {S.}~\bibnamefont
  {Roy}}, \bibinfo {author} {\bibfnamefont {S.}~\bibnamefont {Dietrich}}, \
  and\ \bibinfo {author} {\bibfnamefont {F.}~\bibnamefont {H\"{o}fling}},\
  }\bibfield  {title} {\enquote {\bibinfo {title} {Structure and dynamics of
  binary liquid mixtures near their continuous demixing transitions},}\ }\href
  {\doibase 10.1063/1.4963771} {\bibfield  {journal} {\bibinfo  {journal} {J.
  Chem. Phys}\ }\textbf {\bibinfo {volume} {145}},\ \bibinfo {pages} {134505}
  (\bibinfo {year} {2016})}\BibitemShut {NoStop}%
\bibitem [{\citenamefont {Pathania}, \citenamefont {Chakraborty},\ and\
  \citenamefont {H\"{o}fling}(2021)}]{pathania2021}%
  \BibitemOpen
  \bibfield  {author} {\bibinfo {author} {\bibfnamefont {Y.}~\bibnamefont
  {Pathania}}, \bibinfo {author} {\bibfnamefont {D.}~\bibnamefont
  {Chakraborty}}, \ and\ \bibinfo {author} {\bibfnamefont {F.}~\bibnamefont
  {H\"{o}fling}},\ }\bibfield  {title} {\enquote {\bibinfo {title} {Continuous
  demixing transition of binary liquids: Finite‐size scaling from the
  analysis of sub‐systems},}\ }\href {\doibase 10.1002/adts.202000235}
  {\bibfield  {journal} {\bibinfo  {journal} {Adv. Theory Simul.}\ }\textbf
  {\bibinfo {volume} {4}},\ \bibinfo {pages} {2000235} (\bibinfo {year}
  {2021})}\BibitemShut {NoStop}%
\bibitem [{\citenamefont {Ferrenberg}, \citenamefont {Xu},\ and\ \citenamefont
  {Landau}(2018)}]{ferrenberg2018}%
  \BibitemOpen
  \bibfield  {author} {\bibinfo {author} {\bibfnamefont {A.~M.}\ \bibnamefont
  {Ferrenberg}}, \bibinfo {author} {\bibfnamefont {J.}~\bibnamefont {Xu}}, \
  and\ \bibinfo {author} {\bibfnamefont {D.~P.}\ \bibnamefont {Landau}},\
  }\bibfield  {title} {\enquote {\bibinfo {title} {Pushing the limits of
  {{Monte Carlo}} simulations for the three-dimensional {{Ising}} model},}\
  }\href {\doibase 10.1103/PhysRevE.97.043301} {\bibfield  {journal} {\bibinfo
  {journal} {Phys. Rev. E}\ }\textbf {\bibinfo {volume} {97}},\ \bibinfo
  {pages} {043301} (\bibinfo {year} {2018})}\BibitemShut {NoStop}%
\bibitem [{\citenamefont {Parry}\ and\ \citenamefont
  {Rascón}(2024)}]{parry2024}%
  \BibitemOpen
  \bibfield  {author} {\bibinfo {author} {\bibfnamefont {A.~O.}\ \bibnamefont
  {Parry}}\ and\ \bibinfo {author} {\bibfnamefont {C.}~\bibnamefont
  {Rascón}},\ }\bibfield  {title} {\enquote {\bibinfo {title} {Wetting,
  algebraic curves, and conformal invariance},}\ }\href {\doibase
  10.1103/physrevlett.133.238001} {\bibfield  {journal} {\bibinfo  {journal}
  {Phys. Rev. Lett.}\ }\textbf {\bibinfo {volume} {133}},\ \bibinfo {pages}
  {238001} (\bibinfo {year} {2024})}\BibitemShut {NoStop}%
\bibitem [{\citenamefont {Indekeu}\ and\ \citenamefont
  {Koga}(2022)}]{indekeu2022}%
  \BibitemOpen
  \bibfield  {author} {\bibinfo {author} {\bibfnamefont {J.~O.}\ \bibnamefont
  {Indekeu}}\ and\ \bibinfo {author} {\bibfnamefont {K.}~\bibnamefont {Koga}},\
  }\bibfield  {title} {\enquote {\bibinfo {title} {Wetting and nonwetting near
  a tricritical point},}\ }\href {\doibase 10.1103/physrevlett.129.224501}
  {\bibfield  {journal} {\bibinfo  {journal} {Phys. Rev. Lett.}\ }\textbf
  {\bibinfo {volume} {129}},\ \bibinfo {pages} {224501} (\bibinfo {year}
  {2022})}\BibitemShut {NoStop}%
\bibitem [{Note1()}]{Note1}%
  \BibitemOpen
  \bibinfo {note} {Using $L_2$ regularization during training is a standard
  practice in machine learning which penalizes large values in the neural
  network parameters. This supports the generation of less complex (`smooth')
  models, which is particularly suitable for the functional mapping
  $c_1^{(i)}(\protect \mathbf {r}; [\rho _1, \rho _2], T)$.}\BibitemShut
  {Stop}%
\bibitem [{Zen()}]{Zenodo}%
  \BibitemOpen
  \href@noop {} {}\bibinfo {note} {Code and models available at
  \url{https://github.com/SilasRobitschko/SWNeural}; complete simulation
  dataset available at
  \url{https://doi.org/10.5281/zenodo.15777201}.}\BibitemShut {Stop}%
\bibitem [{\citenamefont {Hermann}\ and\ \citenamefont
  {Schmidt}(2021)}]{hermann2021}%
  \BibitemOpen
  \bibfield  {author} {\bibinfo {author} {\bibfnamefont {S.}~\bibnamefont
  {Hermann}}\ and\ \bibinfo {author} {\bibfnamefont {M.}~\bibnamefont
  {Schmidt}},\ }\bibfield  {title} {\enquote {\bibinfo {title} {Noether’s
  theorem in statistical mechanics},}\ }\href {\doibase
  10.1038/s42005-021-00669-2} {\bibfield  {journal} {\bibinfo  {journal}
  {Commun. Phys.}\ }\textbf {\bibinfo {volume} {4}},\ \bibinfo {pages} {176}
  (\bibinfo {year} {2021})}\BibitemShut {NoStop}%
\bibitem [{\citenamefont {Samm\"{u}ller}\ and\ \citenamefont
  {Schmidt}(2025)}]{sammueller2025mu}%
  \BibitemOpen
  \bibfield  {author} {\bibinfo {author} {\bibfnamefont {F.}~\bibnamefont
  {Samm\"{u}ller}}\ and\ \bibinfo {author} {\bibfnamefont {M.}~\bibnamefont
  {Schmidt}},\ }\href {\doibase 10.48550/ARXIV.2506.15608} {\enquote {\bibinfo
  {title} {Determining the chemical potential via universal density functional
  learning},}\ } (\bibinfo {year} {2025}),\ \Eprint
  {http://arxiv.org/abs/2506.15608} {arXiv:2506.15608 [cond-mat.soft]}
  \BibitemShut {NoStop}%
\bibitem [{\citenamefont {Klink}\ and\ \citenamefont
  {Gross}(2014)}]{klink2014}%
  \BibitemOpen
  \bibfield  {author} {\bibinfo {author} {\bibfnamefont {C.}~\bibnamefont
  {Klink}}\ and\ \bibinfo {author} {\bibfnamefont {J.}~\bibnamefont {Gross}},\
  }\bibfield  {title} {\enquote {\bibinfo {title} {A density functional theory
  for vapor–liquid interfaces of mixtures using the perturbed-chain polar
  statistical associating fluid theory equation of state},}\ }\href {\doibase
  10.1021/ie4029895} {\bibfield  {journal} {\bibinfo  {journal} {Ind. Eng.
  Chem. Res.}\ }\textbf {\bibinfo {volume} {53}},\ \bibinfo {pages}
  {6169--6178} (\bibinfo {year} {2014})}\BibitemShut {NoStop}%
\bibitem [{\citenamefont {Rehner}, \citenamefont {Bardow},\ and\ \citenamefont
  {Gross}(2023)}]{rehner2023}%
  \BibitemOpen
  \bibfield  {author} {\bibinfo {author} {\bibfnamefont {P.}~\bibnamefont
  {Rehner}}, \bibinfo {author} {\bibfnamefont {A.}~\bibnamefont {Bardow}}, \
  and\ \bibinfo {author} {\bibfnamefont {J.}~\bibnamefont {Gross}},\ }\bibfield
   {title} {\enquote {\bibinfo {title} {Modeling mixtures with {PCP-SAFT}:
  Insights from large-scale parametrization and group-contribution method for
  binary interaction parameters},}\ }\href {\doibase
  10.1007/s10765-023-03290-3} {\bibfield  {journal} {\bibinfo  {journal} {Int.
  J. Thermophys.}\ }\textbf {\bibinfo {volume} {44}},\ \bibinfo {pages} {179}
  (\bibinfo {year} {2023})}\BibitemShut {NoStop}%
\bibitem [{\citenamefont {Bursik}\ \emph {et~al.}(2025)\citenamefont {Bursik},
  \citenamefont {Karadimitriou}, \citenamefont {Steeb},\ and\ \citenamefont
  {Gross}}]{bursik2025}%
  \BibitemOpen
  \bibfield  {author} {\bibinfo {author} {\bibfnamefont {B.}~\bibnamefont
  {Bursik}}, \bibinfo {author} {\bibfnamefont {N.}~\bibnamefont
  {Karadimitriou}}, \bibinfo {author} {\bibfnamefont {H.}~\bibnamefont
  {Steeb}}, \ and\ \bibinfo {author} {\bibfnamefont {J.}~\bibnamefont
  {Gross}},\ }\href {\doibase 10.48550/ARXIV.2506.21007} {\enquote {\bibinfo
  {title} {Static contact angles of mixtures: Classical density functional
  theory and experimental investigation},}\ } (\bibinfo {year} {2025}),\
  \Eprint {http://arxiv.org/abs/2506.21007} {arXiv:2506.21007
  [physics.flu-dyn]} \BibitemShut {NoStop}%
\end{thebibliography}%

\end{document}